\documentclass[%
 reprint,
 superscriptaddress,
 amsmath,amssymb,
 aps,
prx,
floatfix,
]{revtex4-2}
\usepackage{lipsum} 
\usepackage{graphicx}
\usepackage{dcolumn}
\usepackage{bm}
\usepackage{nicefrac}
\usepackage{siunitx}
\usepackage[T1]{fontenc}
\usepackage[columnwise]{lineno}
\newcolumntype{Y}[1]{%
  >{\small\raggedright\everypar{\hangindent=1em}\arraybackslash}p{#1}%
}

\usepackage{xcolor}

\usepackage{array}
\usepackage{multirow}
\usepackage{amsmath}

\newcolumntype{R}{>{$}r<{$}}
\newcolumntype{L}{>{$}l<{$}}
\newcolumntype{M}{R@{${}={}$}L}

\usepackage{orcidlink}
\usepackage{pdfpages}
\setboolean{@twoside}{false}
\makeatletter
\AtBeginDocument{\let\LS@rot\@undefined}
\usepackage{hyperref}

\begin{document}
\title{Modular Autonomous Virtualization System\\for Two-Dimensional Semiconductor Quantum Dot Arrays}

\author{Anantha S. Rao\orcidlink{0000-0001-6272-0327}}
\affiliation{Joint Center for Quantum Information and Computer Science, University of Maryland, College Park, Maryland 20742, USA}
\affiliation{Department of Physics, University of Maryland, College Park, Maryland 20742, USA}

\author{Donovan Buterakos\orcidlink{0000-0002-7279-9769}}
\affiliation{Joint Center for Quantum Information and Computer Science, University of Maryland, College Park, Maryland 20742, USA}
\affiliation{National Institute of Standards and Technology, Gaithersburg, Maryland 20899, USA}

\author{Barnaby van Straaten\orcidlink{0009-0008-2113-8523}}
\affiliation{QuTech and Kavli Institute of Nanoscience, Delft University of Technology, P.O. Box 5046, 2600 GA Delft, The Netherlands}

\author{Valentin John}
\affiliation{QuTech and Kavli Institute of Nanoscience, Delft University of Technology, P.O. Box 5046, 2600 GA Delft, The Netherlands}

\author{Cécile X. Yu}
\affiliation{QuTech and Kavli Institute of Nanoscience, Delft University of Technology, P.O. Box 5046, 2600 GA Delft, The Netherlands}

\author{Stefan D. Oosterhout}
\affiliation{QuTech and Netherlands Organisation for Applied Scientific Research (TNO), Delft, The Netherlands}

\author{Lucas Stehouwer}
\affiliation{QuTech and Kavli Institute of Nanoscience, Delft University of Technology, P.O. Box 5046, 2600 GA Delft, The Netherlands}

\author{Giordano Scappucci\orcidlink{0000-0003-2512-0079}}
\affiliation{QuTech and Kavli Institute of Nanoscience, Delft University of Technology, P.O. Box 5046, 2600 GA Delft, The Netherlands}

\author{Menno Veldhorst\orcidlink{0000-0001-9730-3523}}
\affiliation{QuTech and Kavli Institute of Nanoscience, Delft University of Technology, P.O. Box 5046, 2600 GA Delft, The Netherlands}

\author{Francesco Borsoi\orcidlink{0000-0001-9398-7614}}
\altaffiliation[Present address: ]{NNF Quantum Computing Programme, Niels Bohr Institute, University of Copenhagen, Blegdamsvej 17, 2100 Copenhagen, Denmark}
\affiliation{QuTech and Kavli Institute of Nanoscience, Delft University of Technology, P.O. Box 5046, 2600 GA Delft, The Netherlands}

\author{Justyna P. Zwolak\orcidlink{0000-0002-2286-3208}}
\email{Contact author: jpzwolak@nist.gov}
\affiliation{Joint Center for Quantum Information and Computer Science, University of Maryland, College Park, Maryland 20742, USA}
\affiliation{Department of Physics, University of Maryland, College Park, Maryland 20742, USA}
\affiliation{National Institute of Standards and Technology, Gaithersburg, Maryland 20899, USA}

\date{\today}
\begin{abstract}
Arrays of gate-defined semiconductor quantum dots are among the leading candidates for building scalable quantum processors. 
High-fidelity initialization, control, and readout of spin qubit registers require exquisite and targeted control over key Hamiltonian parameters that define the electrostatic environment. 
However, due to the tight gate pitch, capacitive crosstalk between gates hinders independent tuning of chemical potentials and interdot couplings. 
While virtual gates offer a practical solution, determining all the required cross-capacitance matrices accurately and efficiently in large quantum dot registers is an open challenge.
Here, we establish a modular automated virtualization system (MAViS)---a general and modular framework for autonomously constructing a complete stack of multilayer virtual gates in real time. 
Our method employs machine learning techniques to rapidly extract features from two-dimensional charge stability diagrams.  
We then utilize computer vision and regression models to self-consistently determine all relative capacitive couplings necessary for virtualizing plunger and barrier gates in both low- and high-tunnel-coupling regimes.
Using MAViS, we successfully demonstrate accurate virtualization of a dense two-dimensional array comprising ten quantum dots defined in a high-quality Ge/SiGe heterostructure. 
Our work offers an elegant and practical solution for the efficient control of large-scale semiconductor quantum dot systems.
\end{abstract}

\maketitle

\section{Introduction}
Qubits that utilize the spin degree of freedom of charge carriers confined within semiconductor quantum dots (QDs) show great promise for practical large-scale quantum computation~\cite{Burkard21-SSQ}. 
In the past decade, silicon spin qubit systems based on one-dimensional arrays have demonstrated long coherence times, high-fidelity single- and two-qubit gates, coherent quantum information transfer and compatibility with industrial manufacturing techniques~\cite{veldhorst2014addressable, Yoneda2018, Takeda2021, Mills22-TSP, Noiri22-FUG, Philips22-UCS, Zwerver2022, Weinstein2023, Desmet2024, Huang2024, Neyens2024}. 
In more recent years, planar germanium QD hole-spin qubits have emerged as an alternative semiconductor technology that can ease certain challenges in qubit control and device engineering~\cite{Scappucci2021}. 
In particular, the strong spin-orbit coupling and small effective mass of holes in germanium~\cite{Watzinger2018, Lodari2021}, together with a highly uniform and low noise material platform~\cite{Sammak2019, Hendrickx2018, Stehouwer2023}, have sparked tunable QDs~\cite{vanRiggelen2021, Borsoi22-QCA, Hsiao2024} and spin qubits systems arranged in two dimensions~\cite{Hendrickx_2021, Zhang2023, Wang2024}. 
As compelling fabrication methods further advance providing routes for engineering of large QD registers~\cite{Ha22-FDQ, George2024}, it is expected that automation will be playing a pivotal role in orchestrating precise quantum operations at various levels~\cite{Durrer2020, Zwolak20-AQD, Ziegler22-TAR, Zubchenko24-ABQ, Kalantre17-MLD, Zwolak21-RBI, Zwolak21-AAQ, Moon20-ATQ, Ziegler22-TRA, Huang2024}. 

To meet the stringent requirements of high-fidelity operations in spin qubit arrays, targeted manipulation of key Hamiltonian parameters such as chemical potentials and tunnel couplings is required~\cite{Loss98-QCD, Vandersypen17-ISQ, Rimbach2023}.
While dedicated plunger and barrier gates are carefully engineered in QD devices to tune these properties, in practice, metallic gates are capacitively coupled to each other because of their close proximity, causing crosstalk challenges.  
To overcome cross-capacitance effects, the community has adopted \textit{virtual gates}. 
Virtual gates are defined as linear combinations of multiple physical gates, with \textit{virtual matrices} encoding information on how each gate affects a specific array parameter~\cite{Perron15-QSB, Hensgens17-FHQ, Hensgens18-PhD, Volk2019}. 
As the QD registers scale in size and complexity, developing methods that calibrate virtual matrices in an accurate, efficient, and autonomous manner becomes crucial for facilitating high-level and high-fidelity control~\cite{Zwolak21-AAQ}.

In this work, we advance this effort by combining state-of-the-art QD arrays with modern machine learning (ML) to address the open challenge of virtualization. 
We propose and validate a modular automated virtualization system (MAViS), a framework to achieve orthogonal control of the electrostatic potential landscape. 
We test bed the approach on a two-dimensional (2D) germanium ten QD array by defining a multilayer stack of virtual gates that accurately control the QD energies and couplings.
Here, we take advantage of prior works aiming to automate the operation of semiconductor qubits to build a framework that can efficiently fine-tune multi-QD devices.

Initial virtualization methods relied on laboratory heuristics, along with device-specific information processing~\cite{Qiao20-CME, Hsiao20-EOT}.
With the advent of modern machine learning, automated methods relying on ML models and traditional curve fitting to identify orthogonal plunger gates and mitigate capacitative couplings were first proposed in Refs.~\cite{Ziegler23-AEC, Oakes2021automatic}, improving on minimal computer-automated tuning algorithms~\cite{Mills19-CAT, Lapointe-Major19-ATQ}.
The prescribed methods extract the inclinations of transition lines from 2D charge stability diagrams (CSDs) to define the virtualization matrix.
However, as QDs in large arrays decouple from the reservoirs, causing a reduced exchange rate of carriers with respect to the typical measurement scan rates, the 2D CSDs become more challenging to interpret and analyze due to, e.g., secondary charge transitions and more pronounced latching transition lines~\cite{Borsoi22-QCA}.
Our method is articulated on several virtualization layers, exploits both ML and classic analysis techniques for robustness, and has a larger scope than existing approaches, extending beyond plunger gate virtualization. 
By ultimately enabling the definition of virtual barrier gates that tune the interdot couplings over large ranges, without affecting the QD potentials, we build an operation recipe that provides control over the key Hamiltonian parameters defining the electrostatics of the QD system.
Finally, our ability to accurately decode information from complex 2D CSDs enables us to study the boundaries of the linear virtual gates approach and provide a method for using control parameters beyond such limits.  
Our work demonstrates the full virtualization of a dense 2D array of ten QDs in a planar germanium quantum well in the few-hole regime. 
We start with a device hosting weakly coupled QDs pretuned via manual operations and a set of unvirtualized sensor, plunger, and barrier gates. 
By leveraging the MAViS, we then seamlessly transition into a fully orthogonal gate space, achieving complete virtualization in an automated and efficient manner.

The organization of the paper is as follows: 
In Sec.~\ref{sec:methods}, we present an overview of the virtualization framework, including the description of the full virtualization stack in Sec.~\ref{ssec:virt_stack}, the experimental setup used to test the MAViS in Sec.~\ref{ssec:exp_setup}, and the data processing techniques in Sec.~\ref{ssec:image_proc}.
The plunger- and barrier-specific tools are described in Secs.~\ref{ssec:plunger_virt} and~\ref{ssec:barrier_virt}, respectively.
The performance of the MAViS in autonomously defining a set of the virtual plunger and barrier gates to control the ten-QD device is presented in Sec.~\ref{sec:results}.
Finally, in Sec.~\ref{sec:conclusion} we summarize our results and discuss future outlook.

\section{Methods}
\label{sec:methods}
A typical multi-QD spin qubit device is controlled by a set of $N_{\mathrm{P}}$ plunger gates $\mathbf{P}$, $N_{\mathrm{B}}$ barrier gates $\mathbf{B}$, and $N_{\mathrm{S}}$ charge-sensing plunger gates $\mathbf{S}$. 
Tuning the device into a QD regime requires an extensive search over relatively large voltage ranges, with the capacitive crosstalk between gates further complicating the search.
We propose a multilayer virtualization stack that is both device agnostic and modular, as illustrated in Table~\ref{tab:virtual_steps}.

\begin{table*}[t]
\renewcommand{\arraystretch}{1.1}
\renewcommand{\tabcolsep}{2pt}
\caption{\label{tab:virtual_steps}
The five virtualization layers implemented in MAViS.
The first column indicates the layer number.
The second column gives a description of each layer.
Columns three and four give the output and input gates for each layer, respectively.
The equation defining the transformation between the input and output gates is given in the last column.
}
\begin{ruledtabular}
\begin{tabular}{>{\centering\arraybackslash}m{0.9cm}Y{9.5cm}>{\centering\arraybackslash}m{1.95cm}>{\centering\arraybackslash}m{1.65cm}>{\centering\arraybackslash}m{1.3cm}}
{\centering Layer} & 
{Description} & 
{\centering Output gates} &
{\centering Input gates} &
{\centering Trans.} \\ 
\hline
1 & Charge sensor virtualization &  $[\mathbf{vS}, \mathbf{vP}, \mathbf{vB}]$ & $[\mathbf{S}, \mathbf{P}, \mathbf{B}]$ & Eq.~(\ref{eq:sensor_virtualization})\\ 
2 & Plunger gate orthogonalization & $\mathbf{O}$ & $\mathbf{vP}$ & Eq.~(\ref{eq:plunger_orth_pulse}) \\ 
3 & Plunger gate normalization: orthogonalization of plungers with uniform charging voltages & \multirow{2}{*}{$\mathbf{N}$} & \multirow{2}{*}{$\mathbf{O}$} & \multirow{2}{*}{Eq.~(\ref{eq:plunger_norm_pulse})} \\ 
4 & Barrier coarse virtualization: ensuring independent barriers to plungers control in weak-coupling regime & \multirow{2}{*}{$\mathbf{J}$} & \multirow{2}{*}{$[\mathbf{vB}, \mathbf{N}]$} & \multirow{2}{*}{Eq.~(\ref{eq:off_regime_pulse})} \\ 
5 & Barrier fine virtualization: ensuring independent barriers to plungers control in high-coupling regime & \multirow{2}{*}{$\mathbf{K}$} & \multirow{2}{*}{$[\mathbf{J}, \mathbf{N}]$} & \multirow{2}{*}{Eq.~(\ref{eq:on_regime_pulse})} \\ 
\end{tabular}
\end{ruledtabular}
\end{table*}

\vspace{-5pt}
\subsection{Virtualization stack}
\label{ssec:virt_stack}
To ensure the high sensitivity of the charge sensors while the QDs are calibrated, it is convenient to first address the compensation to the charge sensor plunger gates by exploiting a set of virtual gates $\{\mathbf{vP}, \mathbf{vB}\} =\{\mathrm{vP}_i, \mathrm{vB}_j\,|\, i=1,\dots, N_{\mathrm{P}};\, j=1,\dots, N_{\mathrm{B}}\}$ that can be used to maintain the charge sensors tuned to their most sensitive voltage points (layer 1 in Table~\ref{tab:virtual_steps}). 
Achieving this step requires knowledge of the relative lever arms from the plunger and barrier gates controlling the QDs to the plunger gates of the charge sensors ($\mathrm{S}_k$, with $k\in[1, \dots, N_{\mathrm{S}}]$). 
The link between sensor-virtualized gates and real gates is defined through the virtual matrix $M_1$ and reads:
\vspace{-3pt}
\begin{equation}
    \label{eq:sensor_virtualization}
    [\mathbf{vS}, \mathbf{vP}, \mathbf{vB}] = M_1 \cdot [\mathbf{S}, \mathbf{P}, \mathbf{B}],\vspace{-3pt}
\end{equation}
with the absolute value of the entries of the inverse matrix $M^{-1}_1$ being the relative lever arms of the QD plunger gates ($\alpha_{k, N_{\mathrm{S}} + i}^{S}$) and of the barrier gates ($\alpha_{k, N_S + N_{\mathrm{P}} + j}^{S}$) that are extracted experimentally. As an example, if an increase of $1$~\si{\milli\volt} in the barrier $\mathrm{B}_1$ produces a shift of $-0.2$~\si{\milli\volt} to the position of Coulomb peak associated with the third charge sensor in the $\mathrm{S}_3$ space, the entry $(3, N_S + N_{\mathrm{P}}+ 1)$ of $M^{-1}_1$ reads $-0.2$, and $0.2$ is the relative lever arm of $\mathrm{B}_1$ on $\mathrm{S}_3$.  

We also note that splitting the virtualization stack into multiple layers is advantageous for keeping virtual matrices in basic forms.
While $M^{-1}_1$ appears as a large matrix, in practice, it is a diagonal matrix exhibiting off-diagonal elements in only the first $N_{\mathrm{S}}$ rows.
Furthermore, because each of the sensor-virtualized gates $\mathbf{vP}$ and $\mathbf{vB}$ can (by design) be varied without otherwise changing the compensations to the charge sensors, any control parameter built as a linear combination of these virtual gates will also keep all charge sensors tuned.
This allows building up higher-hierarchy virtualization layers using the newly defined $\{\mathbf{vP}, \mathbf{vB}\}$ gates, while the sensor compensations are adjusted automatically as part of the virtual gates.

For spin control and readout, it is practical to operate in an orthogonal plunger gate space where each plunger controls only the designated site's chemical potential independently from the other gates.
This is addressed in layer 2 of MAViS, which we call \textit{plunger gate orthogonalization}, where a set of \textit{ad hoc} virtual plunger gates $\mathbf{O}$ provide orthogonal control over the chemical potentials through a second virtual matrix $M_2$ as
\vspace{-3pt}
\begin{equation}
    \label{eq:plunger_orth_pulse}
    \mathbf{O} = M_2 \cdot \mathbf{vP},\vspace{-3pt}
\end{equation}
with the absolute value of the entry of the inverse matrix $M^{-1}_2$, $\alpha^{O}_{n,i}$, being the relative lever arm of $\mathrm{vP}_i$ to dot $n$. 
This stage can also be extended to integrate compensations for the cross-capacitance effects from the charge sensor to the QDs, if desired.

Layer 3, which we call \textit{plunger gate normalization}, provides a virtual framework spanned by a new set of virtual plunger gates $\mathbf{N}$ with a homogeneous charging voltage over all sites.
This step helps standardize the measurement window size required to build the next layers and is
achieved through a diagonal matrix $M_3$:
\vspace{-3pt}
\begin{equation}
    \label{eq:plunger_norm_pulse}
    \mathbf{N} = M_3 \cdot \mathbf{O},\vspace{-3pt}
\end{equation}
with the entries $\alpha^{N}_{i,i}$ of the inverse matrix $M^{-1}_3$ defined as $d_i/V_D$ with $V_D$ the target charging voltage, and $d_i$ the charging voltage measured along the $\mathrm{O}_i$ voltage space.

To further isolate each QD for stages such as readout and single-qubit gates, we require independent control over the tunnel couplings between neighboring sites~\cite{Burkard21-SSQ}. 
This independence is achieved in layer 4: \textit{barrier coarse virtualization}, where a set of virtual barriers $\mathbf{J}$ is introduced.
As the qubit idle point is typically defined in the uncoupled regime, we perform this calibration in the weak-coupling (OFF) regime where a linear compensation is shown to be sufficient~\cite{Oosterkamp98-MSQ}.
We construct virtual barriers $\mathrm{J}_i$ as a linear combination of sensor-virtualized barrier $\mathrm{vB}_i$ and normalized plungers $\mathrm{N}_j$:
\vspace{-3pt}
\begin{equation}
    \label{eq:off_regime_pulse}
    \mathbf{J} = M_4 \cdot [\mathbf{vB}, \mathbf{N}],\vspace{-3pt}
\end{equation}
where the entry of $M^{-1}_4$ at the position  $\alpha_{N_{\mathrm{B}} + i,j}^\mathrm{OFF}$ encodes the relative shift of the charge state of dot $i$ as the barrier $\mathrm{vB}_j$ voltage is changed.

\begin{figure*}
    \centering
    \includegraphics[width=\linewidth]{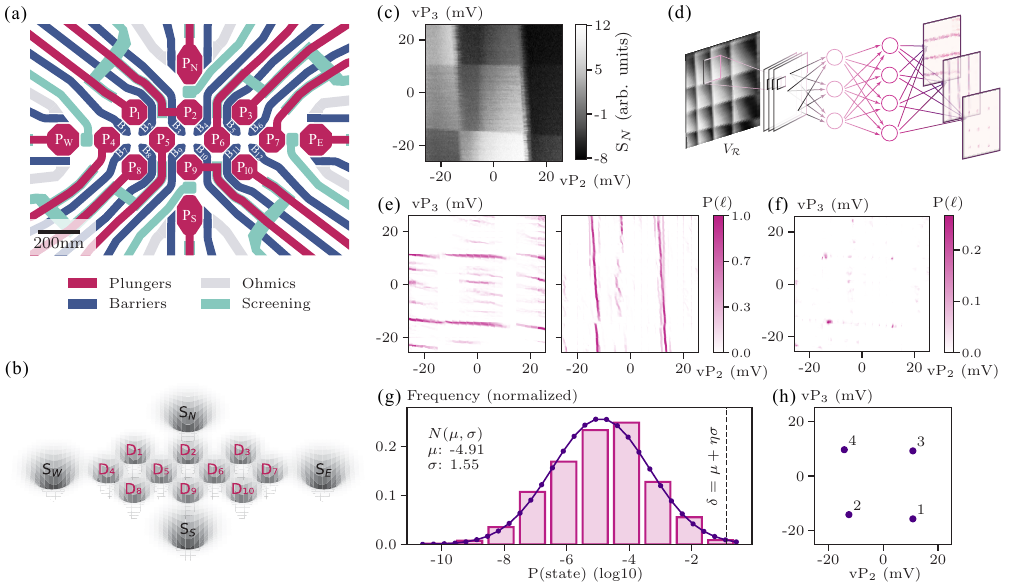}
    \caption{
    Device architecture and virtualization workflow. 
    (a) Layout of a ten-QD array based on a 3-4-3 geometry. 
    Holes are trapped in gate-defined germanium QDs controlled by a set of barrier (dark blue), plunger (magenta), and screening (cyan) gates. 
    (b) Schematic displaying the approximate potential landscape and position of the QDs in the array, with D$_n$, for $n\in[1,\dots,10]$, indicating each dot and S$_N$, S$_E$, S$_S$, and S$_W$ indicating the north, east, south, and west charge sensor, respectively. 
    (c) A typical 2D CSD of a double QD measured via the rf-reflectometry charge sensing on S$_N$. 
    (d) The workflow of the ML model: Each pixel is assigned a probability to be part of a horizontal, vertical, and interdot transition (diagonal and no transition classes not shown). 
    (e) Probability distribution of each pixel to be a horizontal (left) or vertical (right) transition class.
    (f) Probability distribution of each pixel to be an interdot class.
    (g) A Gaussian fit to the log-transformed probability distribution for the interdot class.
    (h) Extracted coordinates of the interdot centers of mass based on dynamic thresholding.}
    \label{fig:overview}
\end{figure*}

The execution of fast two-qubit exchange-based gates requires modulating the tunnel coupling between adjacent QDs across several orders of magnitude~\cite{Loss98-QCD, Hendrickx_2021, Burkard21-SSQ}.
A large voltage pulse on the barrier gates also affects the effective QD positions as the wave functions are brought close to overlap.
This effect is expected to modify the aforementioned capacitive couplings limiting the range of validity of the virtual matrix extracted in the OFF-coupling regime.
As a consequence, barrier control compensations dedicated to the high-coupling (ON) regime are introduced in layer 5.
We anticipate that, depending on the targeted exchange interactions, linear compensations may not suffice and that nonlinear corrections to the plunger gates may be necessary when a barrier is pulsed substantially.
Therefore, we heuristically define quadratically corrected virtual gates $\mathbf{K}$ as follows:
\vspace{-3pt}
\begin{equation}
    \label{eq:on_regime_pulse}
    \mathrm{K}_j = \mathrm{J}_j + \sum_m \left[\alpha^\mathrm{ON}_{j,m} {\mathrm{N}_m}^2 + \beta^\mathrm{ON}_{j,m} \mathrm{N}_m\right],\vspace{-3pt}
\end{equation}
where the $(\sum_m)$ involves only the nearest-neighboring plunger gates to barrier gate $j$, and $\alpha^\mathrm{ON}_{j,m}$, $\beta^\mathrm{ON}_{j,m}$ represent the first and second order compensation coefficients of barrier $J_j$ to the QDs $m$, respectively.

It is important to note that, in general, the charging energy of the QDs varies across different charge states.
This means that the matrices and coefficients in Eqs.~(\ref{eq:sensor_virtualization})--(\ref{eq:on_regime_pulse}) are valid only locally, as full global virtualization across all charge states is not possible without many complex, nonlinear corrections.
However, locally valid virtualization is sufficient for most applications since spin processors are operated in a predefined charge state. 
Thus, it is only necessary to perform virtualization after bringing the device to the desired regime of operation.

\subsection{Experimental setup}
\label{ssec:exp_setup}
The full virtualization approach is demonstrated on a dense array of ten QDs defined in a planar germanium quantum well. 
This device~\cite{Stehouwer24-ESG} is based on low-disorder Ge/SiGe heterostructure grown
on a Ge substrate, with the quantum well separated from the dielectric interface by a $55$-\si{\nano\meter} SiGe barrier~\cite{Stehouwer2023}. 

The device uses a multilayer gate architecture to define an array of ten QDs arranged in a 3-4-3 configuration; see Fig.~\ref{fig:overview}(a). 
Ten plunger gates and 12 barrier gates labeled $\mathrm{P}_i$ and $\mathrm{B}_j$ for $i\in[1, \dots, 10]$ and $j\in[1, \dots, 12]$, respectively, offer control of the array’s electrostatic potential landscape.
Four additional plunger gates, labeled $\mathrm{P}_N$, $\mathrm{P}_E$, $\mathrm{P}_W$, and $\mathrm{P}_S$, control the sensor QD's potentials. 
The ten QDs are labeled D$_n$, with $n\in[1,\dots,10]$, and their four peripheral sensor QDs are labeled S$_N$, S$_E$, S$_W$, and S$_S$, based on their cardinal directions; see Fig.~\ref{fig:overview}(b). 
The additional eight screening gates screen the electric field from the plunger gates to prevent the formation of spurious QDs. 
The screening gates are omitted from the virtualization stack as they are not changed during normal device operation. 

Prior to executing the virtualization stack, the QD array is pretuned to the few-hole regime, with each QD being either in a single-, triple-, or quintuple-hole occupancy for spin qubit manipulation~\cite{John24-TAL}.
The four charge sensors permit fast simultaneous radio-frequency- (rf)-reflectometry charge sensing in combination with video-mode acquisition and frequency multiplexing~\cite{stehlik2015, vigneau2023}.

\subsection{Image processing}
\label{ssec:image_proc}
MAViS has at its basis an ML-based charge transition identifier.
The identification of charge transition lines and interdot transition points (refereed to as \textit{interdots} hereon) from small 2D CSD of different plunger gate voltages, such as shown in Fig.~\ref{fig:overview}(c), requires careful determination of the corners of a honeycomb structure within imperfect images.
This is, in essence, an edge detection and classification problem, a task suitable for ML.

To extract from the experimentally acquired 2D CSDs information useful for automation, we use an ensemble of five convolutional neural network pixel classifiers~\cite{Chollet20-ISU} trained to distinguish each pixel as belonging to five different classes: horizontal, vertical, or diagonal transitions (corresponding to QD formed under left, right, or simultaneously under both plunger gates), interdots, and points where no transitions are present~\cite{Ziegler23-AEC}. 
The pixel classifiers, trained using exclusively simulated data~\cite{Zwolak18-QLD, qf-data}, take as an input a small 2D plunger-plunger voltage scan obtained using a charge sensor, as shown in Fig.~\ref{fig:overview}(c).
It then labels each pixel within the scan with a 5D vector indicating the probability of belonging to each of the five possible charge transition classes.

Prior to running the pixel classifiers, the measured data are preprocessed.
This involves a series of steps.
First, the data are convolved with a Gaussian filter, and the gradient is taken.
The resulting image is then mean normalized, and the normalized gradient is cropped to produce a $32 \times 32$ pixel subimage.

 \begin{figure*}
    \centering
    \includegraphics[width=\linewidth]{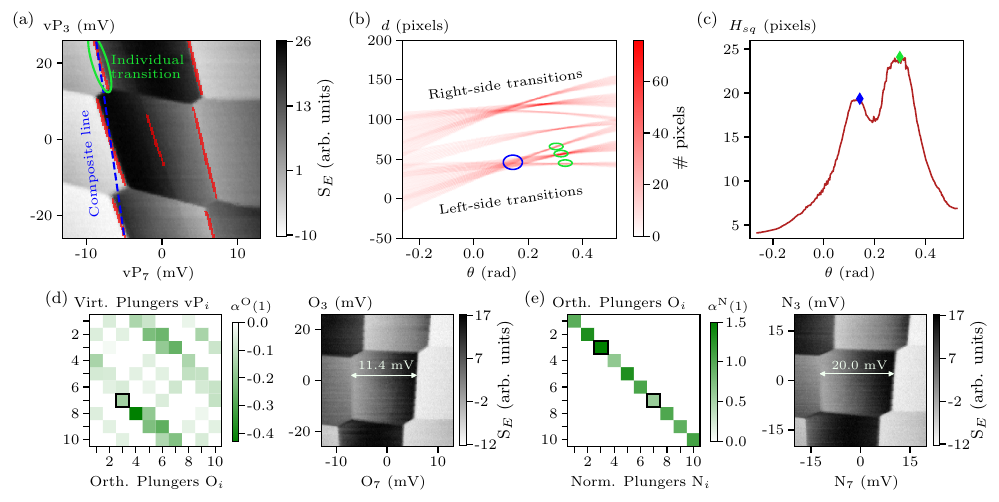}
    \caption{Plunger gates orthogonalization and normalization.
    (a)~Exemplary charge stability diagram spanned by the sensor-virtualized gates $\mathrm{vP}_7$ and $\mathrm{vP}_3$.
    In red, we overlay the output of the pixel classifier module for vertical transitions. 
    The region circled in green indicates a single transition segment, while the blue dashed line indicates the composite line averaged over the three segments of the D$_7$ transition line on the left-hand side of the image. 
    The red line in the center of the image has been erroneously marked by the classifier.
    (b)~The Hough transform of the ML model output. 
    The individual transitions become thin bands, which overlap at a single point corresponding to the composite line (marked in blue).
    The three additional peaks (marked in green) correspond to the individual transitions on the left-hand side. 
    Each pixel has a width of $0.17$~\si{\milli\volt} and a height of $0.35$~\si{\milli\volt}.
    (c)~The sum of the squares of the Hough transform, forming a bimodal distribution corresponding to the composite and individual transitions.
    (d)~Inverse of the virtual matrix $M_2$ obtained from the Hough transforms (left) and a 2D CSD acquired with orthogonalized virtual plunger gates (right).
    (e)~Inverse of the normalized virtual plunger matrix $M_3$ (left) and 2D CSD acquired with normalized virtual plunger gates. 
    In the space spanned by $\mathrm{N}_7$ and $\mathrm{N}_3$, the charging voltages for the specific charge state are constant (20~\si{\milli\volt}), allowing for uniform $x$ and $y$ axes.}
    \label{fig:plunger_virtualization}
\end{figure*}

Each preprocessed image is passed through the ensemble of pixel classifiers in a sliding-window fashion~\cite{Giusti13-FIS, Gouk14-FSW}, and the resulting probability vectors are averaged across the five models.
Since the pixel classifiers sample overlapping windows, the predicted probabilities are averaged over all subimages that contain a given pixel, producing a single image for each class. The individual images codify the likelihood that a horizontal or vertical transition, see Fig.~\ref{fig:overview}(e), or an interdot, see Fig.~\ref{fig:overview}(f), exists at any particular point~\footnote{By design, the pixel classifiers return five-class output. However, the diagonal transition and no transition classes are not relevant to the analysis.}.

To identify the coordinates of the interdots, we implement a dynamic thresholding algorithm that selects the interdots based on the probabilities $\mathrm{P}(\ell)$ given by the ML module, where $\ell$ is the predicted class.
The threshold $\delta$ is determined based on the mean $(\mu)$ and variance $(\sigma^2)$ of a Gaussian fit to the log-transformed interdot class probability distribution, 
\vspace{-3pt}
\begin{equation}
    \label{eq:threshold}
    \delta = \mu+ \eta\sigma,\vspace{-3pt}
\end{equation}
where $\eta$ is set dynamically in the range of $(2.5, 4)$ in order to identify at least the expected number of interdots; see Fig.~\ref{fig:overview}(f).
In the end, a pixel is classified as belonging to an interdot if $\mathrm{P}(\ell)>\delta$.
The final interdot coordinates in the 2D CSD are determined based on the center of mass of each cluster of points classified as an interdot, as illustrated in Fig.~\ref{fig:overview}(g).

\subsection{Plunger virtualization}
\label{ssec:plunger_virt}
A typical 2D CSD acquired as a function of two sensor-virtualized plunger gates exhibits a clearly visible honeycomb pattern, as shown in Fig.~\ref{fig:plunger_virtualization}(a). 
The significant deviation of the vertical and horizontal transition lines from the desirable, perfectly orthogonal pattern indicates the presence of moderate plunger gate-to-dot cross-capacitance and dot-to-dot capacitance that hinder individual control of the chemical potential of each site.
To define a virtual plunger gate framework that allows individual QD control, we perform a Hough transform on the ML model output to find the slopes and locations of the transitions.

From a 2D CSD, the ML module identifies regions that correspond to horizontal and vertical transitions.
For example, in Fig.~\ref{fig:plunger_virtualization}(a), all pixels identified as belonging to vertical transitions are marked in red.
This includes three left-hand side transitions and three right-hand side transitions, as well as an erroneously identified line in the center of the honeycomb.
The ML output is then processed using the Hough transform, resulting in a representation where each transition becomes a thin band in the Hough space, with $\theta$ and $d$ corresponding to the angle and distance from the origin of the transition; see Fig.~\ref{fig:plunger_virtualization}(b).

The peaks in the Hough transform of the ML output correspond to the angles and locations of the transition lines in the original data.
To identify the average slopes of the transitions, all distances in the Hough transformation are squared and summed:
\vspace{-3pt}
\begin{equation}
    \label{eq:hough}
    H_{\text{sq}}(\theta) = \sum_d H(\theta,d)^2,\vspace{-3pt}
\end{equation}
where $H(\theta,d)$ is the Hough transform of the image.
The resulting distribution $H_\text{sq}(\theta)$ will then have a peak at the angle corresponding to the slope of the lines in the image.

The 2D CSDs often contain several nearly collinear transition lines that have slight offset relative to each other due to the presence of interdot transition, as shown in Fig.~\ref{fig:plunger_virtualization}(a).
This effect manifests itself visually in the Hough transformation as several small peaks, shown in Fig.~\ref{fig:plunger_virtualization}(b) with green ovals, each corresponding to an individual transition.
The large peak, shown in Fig.~\ref{fig:plunger_virtualization}(b) with a blue oval, corresponds to a composite line that passes through the centers of the nearly collinear transitions [indicated as a dashed blue line in Fig.~\ref{fig:plunger_virtualization}(a)].
The ability to distinguish between these two cases is essential, as the slope of the composite line will be different from the slopes of the individual transitions themselves, and this difference will be magnified as the size of the interdots increases.

When this occurs, $H_{\text{sq}}(\theta)$ will have a bimodal distribution, with the peak closest to $\theta=\pi/2$ ($\theta=0$) for the case of horizontal (vertical) transitions corresponding to the composite line, and the other peak reflecting the slope of the individual transition segments, as shown in Fig.~\ref{fig:plunger_virtualization} (c).
Our implementation of MAViS is designed to prefer the peak corresponding to individual transitions so that the horizontal and vertical transitions are orthogonal in the final virtualized space, although in principle the composite line can be chosen instead.
After finding the angle $\theta$ corresponding to the individual transitions, the slope of these transitions is then obtained via basic trigonometric functions.
This process is repeated for all nearest-neighbor and next-nearest-neighbor plunger-plunger pairs, yielding the elements of the matrix $M_2^{-1}$ in Eq.~(\ref{eq:plunger_orth_pulse}).

After orthogonalization, the measurement is repeated and the normalization coefficients are found by performing a Hough transform on the new data and finding the difference in $d$ between the largest peak on each side of the image.
This is essentially finding the distance between neighboring transitions, as illustrated in Fig.~\ref{fig:plunger_virtualization}(d).
After finding such distances in all the relevant plunger-plunger maps, the scale factors required to adjust the distances as desired are computed and the median values yield the elements of $M^{-1}_3$. 

\vspace{-5pt}
\subsection{Barrier virtualization}
\label{ssec:barrier_virt}
The next layer of MAViS is the virtualization of the barrier gates with respect to the QD levels.  
Because of the barrier-to-QD capacitive couplings, nonvirtualized barriers result in a shift of the charge state within a CSD as the barrier gate is modified. 
Virtualized barriers are built to ensure that such a charge state, enclosed by four interdots, remains at the charge symmetry point despite varying the tunnel coupling between neighboring QDs.

Barrier virtualization involves determining the virtualization coefficients in Eqs.~(\ref{eq:off_regime_pulse}) and~(\ref{eq:on_regime_pulse}) that will correct for the shifts of the interdot positions as the barrier gates are adjusted. 
The same ML models trained on 2D CSDs obtained from plunger-plunger sweeps are leveraged for this task.
First, the interdot locations in the 2D CSD are determined as described in Sec.~\ref{ssec:image_proc}.
To mitigate errors in tracking individual interdots, the center of the honeycomb formed by a set of four interdots is tracked instead.
If the pixel classifier fails to identify one of the tracked interdots in one of the scans, coordinates for the corresponding interdot from the previous scan are used instead.
Then, we derive the rate at which the charge state shifts as the barrier voltage is changed using one of two methods.

For the OFF regime, the correction coefficients are determined based on the local change in interdot positions.
The local tracking approach involves computing the shift in interdot coordinates only between consecutive scans. 
The average shift of the CSD is defined as the mean of the resulting distribution of pairwise distances normalized by the barrier voltage step.
In addition to being more robust against missing points and boundary effects (e.g., an interdot shifting from the region captured in the CSD), it is also straightforward to identify and exclude outliers caused by mistakes in the pixel classification process (false positive interdot detections).
Local tracking is also more suitable in situations where only a few CSDs are available.

However, local tracking is appropriate only for finding the linear fit coefficient. 
For the quadratic fit observed in the ON regime, a global tracking method is implemented. 
The global method involves direct tracking of the trajectories of the interdot transition points as the barrier gate is pulsed over multiple scans.
The center coordinates as a function of the corresponding barrier voltages are then fit to a linear or quadratic curve, as desired.
While missing or misclassified points can negatively affect the quality of the fit, virtualization in the OFF regime ensures less drastic shifts in the ON regime, making the fit less prone to errors.
Moreover, with a limited number of points involved in the fit, a single misidentified point can substantially affect the resulting fit and lead to a miscalibration, which makes this approach less robust with noisy data.
To overcome this limitation and ensure a high-quality fit, the virtualization in the ON regime is established based on a significantly denser sampling of 2D CSD for every pair of gates.
Since only the neighboring gates need to be virtualized at this step, a larger number of measurements per gate pair can be afforded compared to the OFF regime virtualization.

\section{Results}
\label{sec:results}
\begin{figure*}[!ht]
    \centering
    \includegraphics[width=\linewidth]{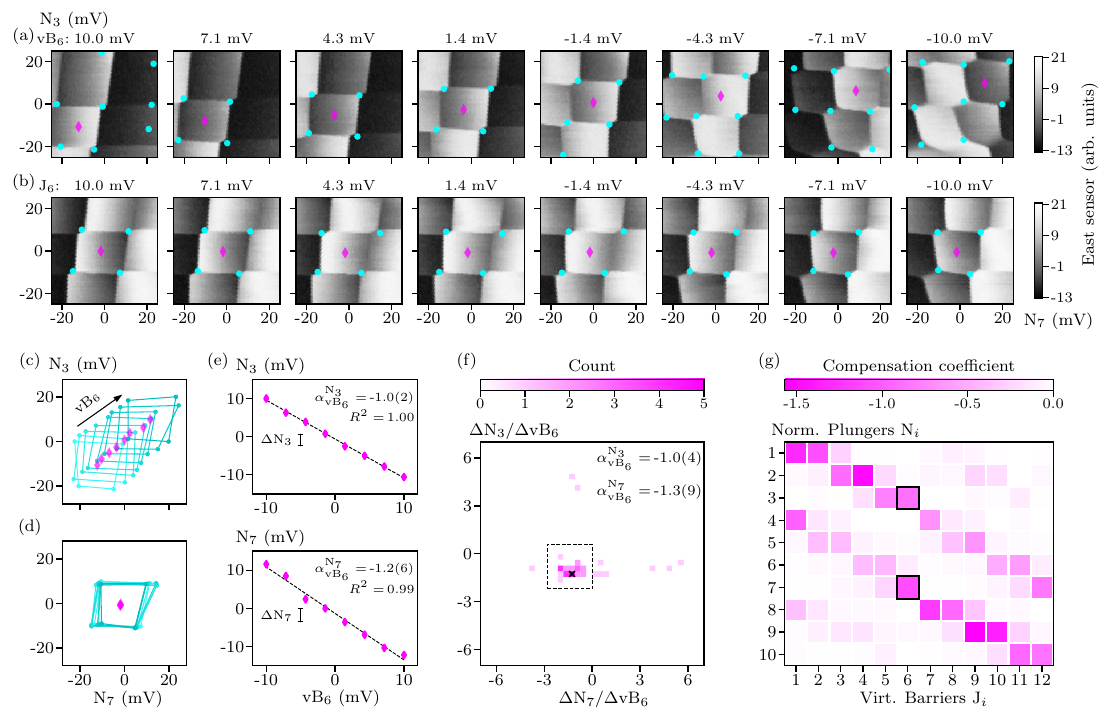}
    \caption{
    Barrier coarse virtualization (OFF regime).
    (a)~Sequence of N$_3$ vs N$_7$ CSDs as a function of barrier v$\mathrm{B}_6$ stepped in the range $[-10, 10]$~\si{\milli\volt} with respect to the starting dc voltage.
    (b) Sequence of N$_3$ vs N$_7$ CSDs as a function of the virtualized barrier J$_6$. 
    The cyan points in (a) and (b) indicate the positions of the interdots returned by the ML module and the magenta diamonds indicate the center of the tracked honeycomb.
    (c), (d) Concise presentations of the evolution of the charge sector (interdots and center of the diamond) extracted from (a) and (b), respectively, showing the effective virtualization of J$_6$, which when varied maintains a constant charge state.
    (e) The fit to the center of the honeycomb positions for plunger gates N$_3$ and N$_7$ as a function of barrier gate v$\mathrm{B}_6$ showing a linear dependence. 
    (f) The distribution of the pairwise distances between all identified interdots between consecutive 2D CSD. 
    The center of this distribution provides the rate of change of interdot position with barrier strength, i.e., the cross-capacitances.
    The dashed box encloses points used to determine the crosstalk coefficients. 
    Points lying outside of this box are considered outliers.
    (g) The resulting capacitive-crosstalk matrix with N$_3$ vs J$_6$ and N$_7$ vs J$_6$ highlighted.}
    \label{fig:barrier-off-regime}
\end{figure*} 

MAViS is designed to autonomously virtualize a set of sensors, plungers, and barrier gates used to define QDs.
It operates through five distinct modules, each corresponding to a specific virtualization layer outlined in Table~\ref{tab:virtual_steps}.
At its core is an ML pixel classifier that extracts high-level feature representations from experimental data.
MAViS leverages horizontal and vertical transition classes to determine the plunger-plunger virtualization coefficients, while the interdot class is used to identify and track interdot transitions for barrier virtualization.
The performance of each virtualization layer is discussed in the remainder of this section, while we focus on the scalability and time requirements of MAViS in Supplemental Material~\cite{supp}.

\subsection{Charge sensors compensation}
\label{ssec:sc_comp}
When searching for clear honeycomblike patterns in 2D CSDs measured via charge sensing, it is critical to maintain the rf-reflectometry charge sensors tuned to their maximum-sensitivity point, i.e., at the flank of one of their Coulomb peaks, throughout the whole voltage scan. 
The high sensitivity of charge sensing is ensured by calibrating the relative lever arm of each gate controlling the QD array to the charge sensors' plunger gates. 
This step does not entail using ML algorithms and exploits only traditional analysis routines to track the charge sensors' Coulomb peak position as a function of each barrier and plunger gate. 

The charge sensor compensation matrix $M_1$ is determined in the first virtualization layer by obtaining the crosstalk of each gate to the four charge sensors.
This is achieved by extracting the slope of the Coulomb peak position as a function of each gate.
The slope of the linear shift in the $\mathbf{S}$ vs $\mathbf{P}$, and $\mathbf{S}$ vs $\mathbf{B}$ space is extracted and fed into the matrix $M^{-1}_1$, as shown in Supplemental Material~\cite{supp}. 
As we move on to the next virtualization layers, only the sensor-virtualized $\mathbf{vP}$ and $\mathbf{vB}$ parameters are considered.

\subsection{Plunger gates virtualization}
\label{ssec:pg_comp}
The plunger gate orthogonalization matrix $M_2$ is determined using the slope extraction method described in Sec.~\ref{ssec:plunger_virt}.
First, a 2D CSD acquired for each pair of plunger gates ($\mathrm{vP}_i$, $\mathrm{vP}_j$), with $i, j \in [1,\dots,10]$, and $i \neq j$, is processed by the ML module to detect the horizontal and vertical charge transition lines; see Fig.~\ref{fig:plunger_virtualization}. 
Then, a Hough transform method is used to extract the slopes of all relevant charge transitions; see Fig.~\ref{fig:plunger_virtualization}(b).
The inverses of the slopes define the elements of the orthogonalization matrix; see the right-hand panel in Fig.~\ref{fig:plunger_virtualization}(d).
When scanning the newly defined virtual plunger gates ($\mathrm{O}_i$, $\mathrm{O}_j$), with $i, j \in [1,\dots,10]$, the honeycomb diagrams display orthogonal transition lines confirming the individual control of each plunger to the corresponding QD, as shown in the left-hand panel in Fig.~\ref{fig:plunger_virtualization}(d).

The goal of the plunger normalization layer is to ensure uniformity of charging voltages.
This step gives three practical advantages.
First, it allows the homogenization of the measurement window required for the next steps. 
Second, it enables us to define all detuning ($\varepsilon_{i,j}$, with $i,j\in [1,\dots,10]$) and total-energy ($U_{i,j}$, with $i,j\in[1,\dots,10]$) axes of all double-QD pairs, often adopted for readout and initialization schemes, with the same simple $45^\circ$-rotation matrix.
Third, when a sizable charge jump occurs, visualizing a charge state with a different charging voltage (due to the few-hole filling structure) immediately triggers the experimentalist or a surveying algorithm to perform a more in-depth verification of the charge state.

\begin{figure*}
    \centering
    \includegraphics[width=\linewidth]{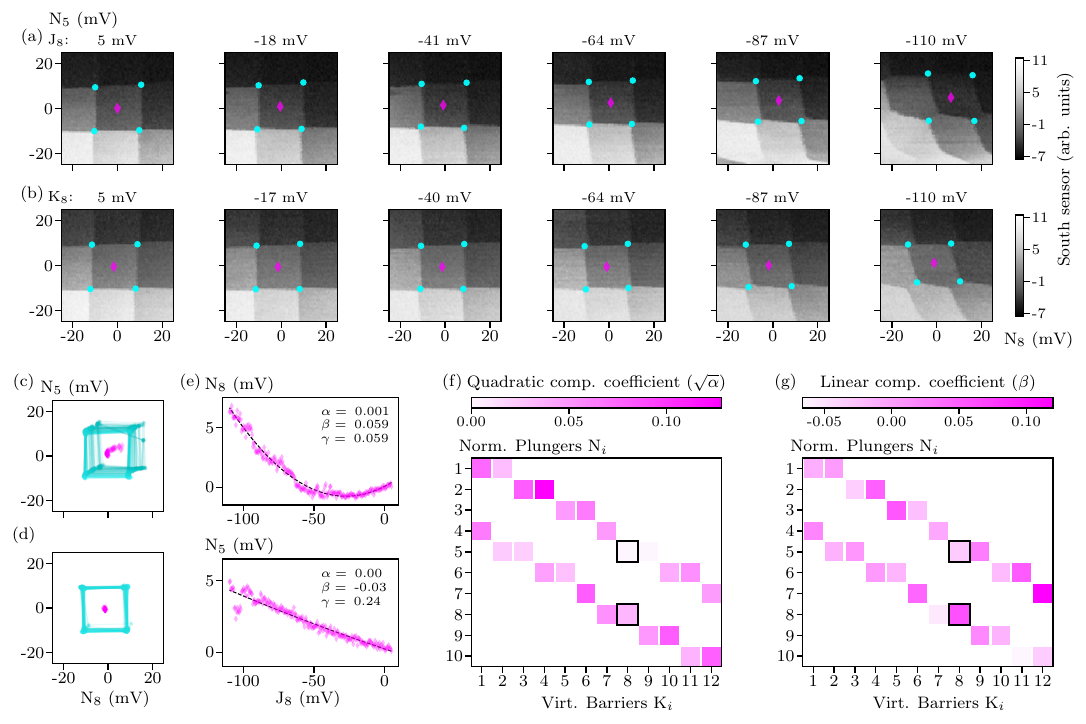}
    \caption{
    Barrier fine virtualization (ON regime).
    (a) Sequence of CSDs N$_5$ vs N$_8$ as a function of J$_8$ in the range $[-110, 5]$~\si{\milli\volt} with respect to the dc reference point.
    (b) Sequence of CSDs N$_5$ vs N$_8$ as a function of K$_8$ in the same range. 
    The cyan points in (a) and (b) indicate the position of the interdots returned by the ML module and the magenta diamonds indicate the center of the tracked honeycomb.
    The shift in the (N$_8$, N$_5$) space of the central charge sector in (a) reveals imperfect virtualization over the large voltage range (115~\si{\milli\volt}). 
    Panel (b), where the finely virtualized gate K$_8$ is adopted, shows a decreased susceptibility to the barrier gate voltage change.
    (c), (d) Concise presentation of the evolution of the charge state while stepping J$_8$ and K$_8$, respectively. 
    The finely calibrated K$_8$ preserves the position of the charge state. 
    (e) The fit to the center of the honeycomb positions for plunger gates N$_5$ and N$_8$ as a function of barrier gate J$_8$ indicating a beyond-linear dependence for N$_5$.
    (f), (g) The quadratic coefficient $\sqrt{\alpha}$ and linear compensation coefficient $\beta$, respectively, with K$_8$ vs N$_5$ and vs N$_8$ highlighted.}
    \label{fig:barrier-on-regime}
\end{figure*}

The uniformity of charging voltages with a target constant spacing between consecutive transitions 
in a new virtual gate space spanned by the normalized plunger gates $\mathrm{N}_i$, with $i\in[1,\dots,10]$, 
is achieved through a diagonal virtualization matrix $M_3$.
The desired spacing across the charge transitions and interdots encompassing the central charge state of interest is set to $V_D=20$~\si{\milli\volt}.
The elements of matrix $M_3$ are determined using the slope extraction method described in Sec.~\ref{ssec:plunger_virt} applied to 2D CSDs acquired after the first plunger orthogonalization is completed.
Since the uniformity of charging voltages is determined in an already orthogonalized plunger space $\mathrm{O}_i$, with $i\in[1,\dots,10]$, the resulting matrix $M_3$ is diagonal; see the right-hand panel in Fig.~\ref{fig:plunger_virtualization}(d) for $M^{-1}_3$.
The homogenized spacing between charge transitions in the ($\mathrm{N}_i$, $\mathrm{N}_j$) space, with $i, j \in [1,\dots,10]$,  is illustrated in the right-hand panel of Fig.~\ref{fig:plunger_virtualization}(e).

\subsection{Barrier virtualization in the OFF-coupling regime}
\label{ssec:bg_off_comp}
Once the plunger gates are orthogonalized and normalized, MAViS proceeds to virtualize the barrier gates to achieve individual control of tunnel couplings without affecting the chemical potential of any QD.
The barrier-plunger capacitive coupling manifests as a change in interdot position in a 2D CSD as the plungers are swept for different barrier gate voltages.
As the barrier v$\mathrm{B}_6$ is stepped from $-10$~\si{\milli\volt} to $10$~\si{\milli\volt}, there is a clearly visible charge state shift across 2D CSDs due to the coupling of the barrier gates to the QD levels is shown in Fig.~\ref{fig:barrier-off-regime}(a).

Extraction of barrier-dot coupling involves choosing for each barrier gate v$\mathrm{B}_i$, for $i\in[1,\dots,12]$, a pair of neighboring virtual plunger gates (N$_i$, N$_j$), for some $i,j\in[1,\dots,10]$, measuring several 2D CSDs while uniformly stepping v$\mathrm{B}_i$, and extracting the shifts in the interdot positions.
For each 2D CSD, the positions of interdots forming the honeycomb of interest and of the corresponding center of the honeycomb are determined using image processing methods described in Sec.~\ref{ssec:image_proc}.
The trajectory of the honeycombs and their centers, shown in Figs.~\ref{fig:barrier-off-regime}(c) and~\ref{fig:barrier-off-regime}(e), suggest a roughly linear relationship between the plunger and barrier gates, as expected due to the low coupling in the OFF regime.

However, on occasion, the pixel classifier fails to identify interdots or identifies false positives [for examples of each, see the first and last 2D CSD in Fig.~\ref{fig:barrier-off-regime}(a)].
Moreover, as the 2D CSD shifts due to crosstalk, interdots may shift into or out of the frame as the barrier voltage is changed.
In such cases, the quality of the linear fit will be significantly affected, resulting in suboptimal accuracy of the extracted slope.
Thus, rather than finding the slopes directly, we rely on the local change in the interdots' position, as described in Sec.~\ref{ssec:barrier_virt}.
 
A histogram of the computed pairwise distances between interdots in consecutive 2D CSDs along each plunger axis is shown in Fig.~\ref{fig:barrier-off-regime}(f).
This frequency distribution of distances allows us to determine the average shift over all interdots, indicated by a black cross in Fig.~\ref{fig:barrier-off-regime}(f).
The outliers due to boundary effects and classifier errors are easily identifiable as stray points far away from the central cluster of points highlighted with a dashed box in Fig.~\ref{fig:barrier-off-regime}(f).

The barrier virtualization coefficients are derived directly from the histogram, as described in Sec.~\ref{ssec:virt_stack}. 
A section of the matrix $M^{-1}_4$, containing all the cross-capacitance compensations, is depicted in Fig.~\ref{fig:barrier-off-regime}(g).
Using the newly defined virtual barrier gates $\mathrm{J}_1, \dots, \mathrm{J}_{12}$, we iterate the procedure for an additional three rounds to refine the coefficients of the matrix.
Our results show that repeating this procedure for an additional two rounds is beneficial and leads to an overall improvement in the accuracy of the estimated parameters (see Supplemental Material~\cite{supp}).
When exploiting the set of virtual barriers $\mathrm{J}_1, \dots, \mathrm{J}_{12}$, the honeycomb diagram shown in 2D CSDs remains fixed with respect to barrier voltage changes as shown in Fig.~\ref{fig:barrier-off-regime}(b). 
The trajectory of the honeycombs and their centers, shown in Fig.~\ref{fig:barrier-off-regime}(d), further validate the derived virtual barrier gates.

\subsection{Barrier virtualization in the ON-coupling regime}
\label{ssec:bg_on_comp}
Similar to the OFF regime, the interdot detection algorithm is deployed to locate and track the honeycomb of interests in the ON regime. 
However, in the ON regime, the capacitive coupling manifests not only as a shift in the trajectory of the center of the honeycombs but also as a change in the honeycomb shape, as visible in Fig.~\ref{fig:barrier-on-regime}(a) and in the simulations in Appendix~\ref{app:simulations}.
Modulating the barrier voltage to achieve the required large coupling strengths affects the effective QD positions as the wave functions are brought close to overlap.
This effect, in turn, modifies the capacitive couplings, limiting the range of validity of the virtual matrix extracted in the OFF-coupling regime.
Thus, the virtual barrier gates constructed in the OFF regime need to be corrected for the strongly coupled ON regime.
Importantly, in the ON regime, deviations from a purely linear trend emerge.
 
In order to find the nonlinear corrections, we use the global tracking method presented in Sec.~\ref{ssec:barrier_virt}.
We note that, in general, this method is more sensitive to errors in the interdot-identification process, such as misidentified interdots, false positives, and boundary effects, as it does not include a simple way of excluding outliers.
However, because we have already performed virtualization in the OFF regime, the large-scale shifts have already been accounted for, which makes boundary effects much less problematic.
For instance, the total shift in Fig.~\ref{fig:barrier-on-regime}(a) is much smaller than the corresponding shift in Fig.~\ref{fig:barrier-off-regime}(a), despite the barrier being sampled over a much larger range, at about $5$~\si{\milli\volt} in the ON regime compared to around $20$~\si{\milli\volt} for the OFF regime.

As the barrier is pulsed and the QDs are brought close together, the center of the honeycomb shifts.
Particularly, as shown in Figs.~\ref{fig:barrier-on-regime}(c) and~\ref{fig:barrier-on-regime}(e), the center of the (N$_5$, N$_8$) 2D CSD is seen to follow a quadratic trajectory as barrier J$_8$ is lowered.
We observe the same trend across all double QDs.
The quadratic and linear compensation coefficients, depicted in Figs.~\ref{fig:barrier-on-regime}(f) and~\ref{fig:barrier-on-regime}(g), respectively, are derived as described in Sec.~\ref{ssec:virt_stack}. 
We observe that corrections to the barrier matrix defined in the ON-coupling regime can be of up to $\sim10$~\% in the first order.
Finally, we note that the square root of the quadratic and linear coefficients are of the same order, and both terms are equally important to virtualizing the barriers. 

Having obtained the best-fit coefficients, we then perform barrier pulses that incorporate quadratic and linear compensations on the plungers, by constructing virtualized barriers, K$_i$ for $i\in[1,\dots,12]$, according to Eq.~(\ref{eq:on_regime_pulse}). 
As the virtual gate K$_i$ is varied, we observe that the map is maintained at the charge symmetry point, allowing for a wide range of barrier pulses, as shown in Fig.~\ref{fig:barrier-on-regime}(b). The compensation matrix values for the linear and quadratic terms, together with measurements on additional double quantum dots, are provided in Supplemental Material~\cite{supp}.

\vspace{-5pt}
\section{Summary and outlook}
\label{sec:conclusion}
Despite significant progress in operating and automating the control of QD devices, gate virtualization remains a challenging and time-consuming process that hinders high-level control of spin qubit arrays. 
In this work, we have introduced MAViS, a modular and scalable framework that enables the autonomous construction of a complete stack of multilayer virtual gates starting from raw plunger and barrier gate voltages and sensor readouts in arbitrary architectures. 
We benchmark MAViS on a 10-QD array with 2D connectivity and demonstrate that full virtualization can be achieved in only $\approx 5$~\si{\hour}, inclusive of both data acquisition and processing (see Supplemental Material~\cite{supp}). 

Motivated by the need for precise control over key Hamiltonian parameters in spin qubit arrays, our approach addresses the virtualization challenge by applying modern ML techniques to state-of-the-art QD arrays. 
We envision our approach to be useful also by other platforms such as in Kitaev chains~\cite{Dvir2023,tenHaaf2024, Bordin2024}, in Andreev qubits~\cite{Hays2021, Pita-Vidal2023}, and in topological readout schemes~\cite{Borsoi2020, Aghaee2024}.

We have designed our methodology to be extremely flexible, as evidenced by the modular structure of the virtualization stack.
Such modularity allows for easy integration with other tools and frameworks without relying on the details of the device-specific software.
It also allows for the incorporation of unit testing, since the virtualization matrices can be validated and errors detected at every step of the process.
Moreover, the modular design enables adaptation and advancement of only selected portions of the virtualization process as necessary.

Likewise, we highlight the generalizability of the ML models.
Despite being trained using exclusively simulated, unvirtualized data~\cite{Ziegler22-TRA}, the model ensemble is able to decode features from charge stability diagrams both before and after the plunger orthogonalization and normalization stages.
Additionally, the models are able to correctly classify CSDs from a 2D array of germanium hole qubits, even though its training data were simulated for electrons in a 1D quantum nanowire, thus demonstrating that our tools are device agnostic. 
Moreover, it is important to highlight that the development and deployment of tools based on ML and signal processing (Hough transform, regression models, etc.) has allowed us to track interdots and map the full dependence of the charge states with barrier voltages. 
Furthermore, numerical simulations revealed that the observed beyond-linear dependence can be caused by an increase in capacitive coupling between the barrier gate and the QDs.

While our virtualization flow has succeeded in virtualizing a large array, we note that several potential points of failure remain.
First, it is possible that the CSD itself might be too noisy for the ML model, especially for QDs near the center of the device, which can be more difficult for the sensors to pick up.
In our case, because the device has multiple sensors, we were always able to find a clear CSD with distinct charge transitions.
By running the analysis on the output from all four sensors and selecting the clearest result, we were able to handle cases where some of the sensors gave suboptimal data.
Second, the pixel classifier occasionally misidentified transitions or identified false positives, particularly in cases with large latching effects.
This also occurred where there was a large gradient across the region defined by a specific charge state, such as the central hexagon in Fig.~\ref{fig:plunger_virtualization}(a), where the pixel classifier marked a red line in the center of the image which does not correspond to any transition.
In both cases, we mitigated these errors by carefully designing the postprocessing methods to be robust to such errors.

At present, MAViS enables the virtualization of the barrier gates with respect to the QD charge states.
An end-to-end virtualization stack would require compensating for the effect of each barrier to every interdot exchange coupling. 
This effect can be studied with precision by mapping the exchange coupling as measured via resonant qubit spectroscopy as a function of virtualized barriers in the ON regime~\cite{Hendrickx20-FTL, Philips22-UCS, John24-TAL}. 
Since compensating for exchange coupling requires qubit characterization rather than decoding information from CSDs, it is left for future work. 

The extracted capacitive coupling matrices obtained from our methods contain rich information about the device and can be analyzed further to gain insight into the effective location of the QDs, the disorder landscape, and the general impurity density over each metallic gate. 
This approach will allow the community to autonomously track QD features for calibration of large-scale QD arrays.

\vspace{-7pt}
\begin{acknowledgments}
The authors are grateful to S.~de~Snoo for support in the control software and to M.~Gullans, C.~White, S.~Gandhari, and all the members of the Veldhorst lab for fruitful discussions.
This research was sponsored in part by the U.S. Army Research Office (ARO) under Awards No. W911NF-23-1-0110 and No. W911NF-23-1-0258.
We acknowledge support from the European Union through the IGNITE project with Grant Agreement No. 101069515 and from the Dutch Research Council (NWO) via the National Growth Fund program Quantum Delta NL (Grant No. NGF.1582.22.001).

The views, conclusions, and recommendations contained in this paper are those of the authors and are not necessarily endorsed nor should they be interpreted as representing the official policies, either expressed or implied, of the U.S. Army Research Office (ARO) or the U.S. Government. 
The U.S. Government is authorized to reproduce and distribute reprints for Government purposes notwithstanding any copyright noted herein. 
Any mention of commercial products is for information only; it does not imply recommendation or endorsement by the National Institute of Standards and Technology.

\end{acknowledgments}

\vspace{-5pt}
\subsection{Data Availability Statement}
The data that support the findings of this article are openly available~\cite{virtualization-dataset}.
Complete figure source files are available via data.nist.gov~\cite{mavis-figures}.

\appendix
\section{Capacitative simulations of quadratic behavior}
\label{app:simulations}
The quadratic shift observed in the interdot position with decreasing barrier voltage, as presented in Fig.~\ref{fig:barrier-on-regime}, can be reproduced using a capacitive model, where the capacitive couplings are given gate voltage dependence. 
In particular, we expect that more negative barrier voltages should draw the dots together, leading to an increased dot-dot and barrier-dot capacitive coupling. 
Our simulations show that the increase in barrier voltage as the barrier gate voltage decreases is sufficient to account for the quadratic dependence we observed, while the increase in the dot-dot coupling widens the interdot. 

\begin{figure*}
\centering
\includegraphics[width=1.\linewidth]{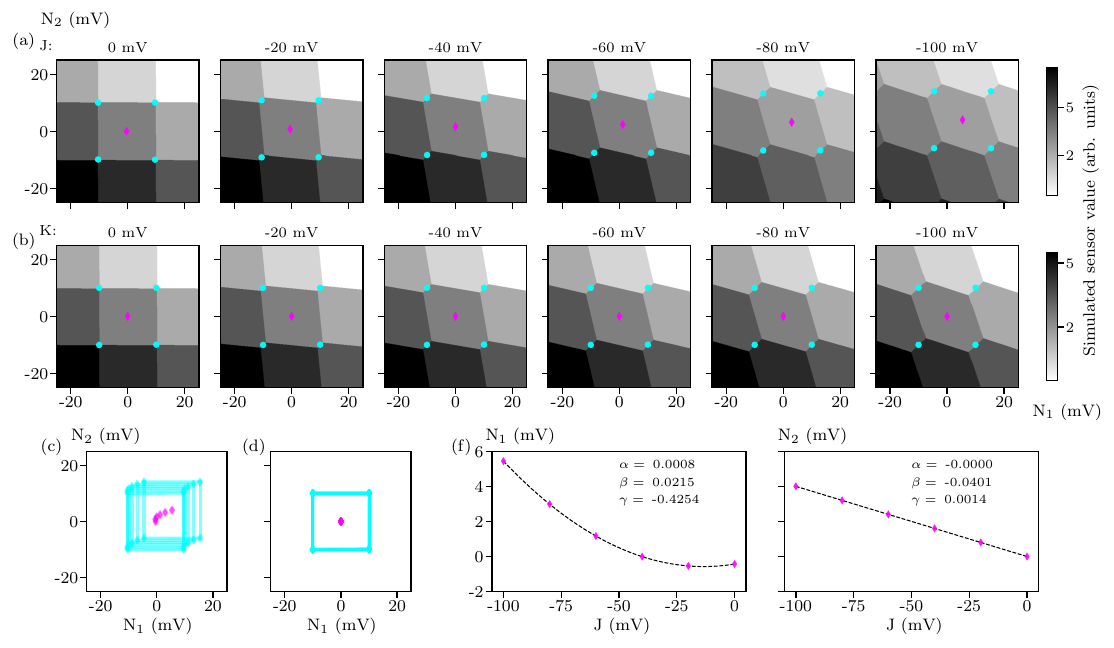}
\caption{
Simulated barrier virtualization in the ON regime, showing the widening of the interdot region and the quadratic shift in the charge state center.
(a), (b) Simulated charge stability diagrams illustrating the shift in interdot transitions (cyan dots) and the center of the $(1, 1)$ charge state (magenta diamonds) as a function of the virtualized barrier in the OFF ($\mathrm{B}$) and ON ($\mathrm{J}$) regimes, respectively.
(c), (d) Overlaid positions of the interdot transitions (cyan) and the [1, 1] charge state center (purple) as functions of $\mathrm{J}$ and $\mathrm{K}$.
(e) The quadratic fit to the position of the $(1, 1)$ charge state center as a function of $\mathrm{J}$ and $\mathrm{K}$.
}
\label{app_figure:simulation}
\end{figure*}

Figure~\ref{app_figure:simulation} shows a simulated recreation of the barrier virtualization into the ON regime, qualitatively reproducing the observed behavior in Fig.~\ref{fig:barrier-on-regime}. 
These simulations were performed using the open-source package QArray~\cite{vanstraaten2024}. 
The system is modeled as a double QD controlled by two plunger gates ($\mathrm{P}_1$, $\mathrm{P}_2$) and one barrier gate ($\mathrm{B}$). 
The capacitive couplings are captured by the following capacitance matrices:
\vspace{-3pt}
\begin{align}
     C_{dd}(\mathrm{B}) &= \begin{bmatrix}
        0 &  0.5 b \\
        0.5 b & 0 
    \end{bmatrix}, \\
     C_{gd}(\mathrm{B}) &=  0.05 \begin{bmatrix}
        1 & 0 & 1 + 0.8 b \\
        0 & 1 & 1
    \end{bmatrix},
\end{align}
where $C_{dd}$ and $C_{gd}$ represent the dot-dot and gate-dot capacitive couplings, respectively, which we allow to depend upon the barrier gate voltage through $b = -\mathrm{B}/10^{3}$. 
We construct the normalized plunger gate voltages, $\mathrm{N}_1$ and $\mathrm{N}_2$, at $\mathrm{B} = 0$~\si{\milli\volt} and define the OFF regime virtualized barrier $\mathrm{J}$ at $\mathrm{B} = -25$~\si{\milli\volt}. 
In the ON regime, the virtualized barrier $\mathrm{K}$ is defined by fitting a quadratic curve to the position of the charge state center as a function of the barrier voltage. 
We find that linear trends, such as shown in Fig.~\ref{fig:barrier-on-regime}(e), are indicative of errors in the linear virtualization coefficients. 
For qualitative agreement with Fig.~\ref{fig:barrier-on-regime}(e), we include an error in the barrier virtualization against $\mathrm{N}_2$, so that the coefficient was taken as $-0.96$ rather than the optimal value of $1.00$. 
This model provides a clear capacitative interpretation of the quadratic trends observed experimentally.

%

\clearpage
\newpage


\section*{Supplementary material}

# Supplemental Material: MAViS: Modular Autonomous Virtualization System
## for Two-Dimensional Semiconductor Quantum Dot Arrays

### Supplementary Section 1.   MAVIS: SCALABILITY AND COMPLEXITY

Our proposed automated virtualization stack (MAViS) is general and applicable to arbitrary quantum dot (QD) device architectures. At its core, MAViS derives the virtual plunger and virtual barrier gates by processing two-dimensional (2D) charge stability diagrams (CSDs) (referred to as "maps" hereon) measured in plunger gates space with the barrier gates adjusted as necessary. Here, we present the complexity analysis of the MAViS framework. We discuss how the number of maps required to virtualize a device scales with the number of metallic gates for one-dimensional (1D) QD arrays before shifting focus to non-1D cases.

### A.   One-dimensional QD arrays

Consider a QD device with $N_S$ charge-sensing plunger gates $\mathbf{S}$, $N_P$ plunger gates $\mathbf{P}$, and $N_B$ barrier gates $\mathbf{B}$. The first layer, charge sensor virtualization, in MAViS requires $(N_S \times N_P + N_S \times N_B)$ maps. Since $N_B \approx N_P + 1$ in most linear arrays, virtualizing sensors requires $O(N_P N_S)$ maps. However, since the magnitude of compensations decays with the distance between the gates, we can avoid measuring gate pairs located farther than a maximum distance. Considering an $n_S$-nearest neighbor approach, where $n_S \ll N_P$ and only the $n = \{1, 2, \dots n_S\}$ nearest neighboring metallic gates in a 1D QD array share a cross-capacitance, the capacitance vanishes with the $(n_S + 1)$ neighbor and we can reduce the number of required maps to $O(n_S N_S) \approx O(N_S)$.

When orthogonalizing plungers (layer 2 in MAViS), a single map suffices to extract the capacitive coupling between two neighboring plungers, which correspond to two coefficients in the $(N_P \times N_P)$ plunger-plunger compensation matrix. Thus, the all-to-all plunger connectivity requires $N_P(N_P - 1)/2 = O(N_P^2)$ maps. However, following the same $n_P$-nearest neighbor approach, the total number of maps for plunger-orthogonalization is $O(N_P)$. Next, to perform normalization (layer 3) we would again require $O(N_P)$ maps.

Barrier virtualization (steps 4 and 5 in MAViS) involves tracking the center of a selected honeycomb in a series of 2D CSDs and fitting a curve to the resulting trajectory. Assuming that each barrier is stepped $m$ times during barrier virtualization, the all-to-all coupling approach would necessitate $[N_P(N_P - 1) \times N_P \times m]/2$ maps. However, since the capacitive coupling of barriers is restricted to the nearest plungers, we only need to worry about the number of (nearest) neighbors of the barriers, $n_B$, that affect the crosstalk while others can be set to zero. In this case, the number of maps required is $[n_B(n_B - 1)N_P m] = O(m N_P)$ both in the low- and high-tunnel coupling regime.

Combining all the steps of MAViS, we obtain a linear scaling for virtualizing sensors and plungers with a constant prefactor given by the number of neighbors with non-vanishing cross-capacitances. For virtualizing barrier gates, we would again require $O(N_P)$ maps with the prefactor given by the number of barrier steps that depend on the experiment and the number of nearest neighbors. Thus, the virtualization framework scales linearly with the number of plungers and barriers for a 1D device.

### B.   Higher-dimensional QD arrays

MAViS would follow the same scaling in higher dimensions, however, the number of nearest neighbors $\{n_S, n_P, n_B\}$ will be dimension dependent. Thus, performing MAViS on a higher dimensional QD device would require $O(n_S \times s)$, $O(n_P \times N_P \times m)$, $O(n_B \times N_B \times m)$ maps for virtualization of sensors, plungers, and barriers, respectively.

In the case of the ten-QD two-dimensional device used in this work, the estimated average time necessary to acquire and process the experimental data with MAViS, as well as the total number of measurements for each virtualization layer, are detailed in Table I. In layer 2, we have only considered the cross capacitance compensations up to the second nearest neighbors. We neglected

**Supplementary Table** I. Modular Automated Virtualization System (MAViS) timing information when applied to the ten-QD device. The estimated mean data acquisition and MAViS processing times are given per map and are calculated based on a total measurement and processing, respectively, time divided by the number of maps in a given layer. For the barrier fine virtualization (layer 5), a large number of maps was acquired to carefully follow the non-linear dependence. However, in practice, only a small fraction of the total number of measurements is required.

| Virt. layer | MAViS step | Data acquisition time [s] | MAViS processing time [s] | Total no. of measurements |
|---|---|---|---|---|
| 1 | Charge sensor virtualization | 0.34 | 0.002 | 88 |
| 2 | Plunger gate orthogonalization | 2.3 | 35.3 | 27 |
| 3 | Plunger gate normalization | 2.4 | 35.3 | 10 |
| 4 | Barrier coarse virtualization | 4.5 | 3.9 | 672 |
| 5 | Barrier fine virtualization | 4.5 | 3.8 | 1440 |

cross capacitances between elements that are very far in the array design. For instance, we assume that the cross capacitance of $P_4$ to $P_2$ and $P_4$ to $P_9$ is negligible. Furthermore, we note that measurements in the barrier fine virtualization step are highly oversampled, with maps acquired for every 0.5 mV shift in the barrier voltage. This was done to enable a highly precise fitting in the high-coupling regime. In practice, $O(10)$ samples are sufficient to characterize the non-linear trend and only maps between neighboring plungers and barriers suffice for barrier virtualization in the low-coupling regime. Taking all of this into account, the total time MAViS would require to virtualize the ten-dot device would be about 2 h 15 min $(8,064$ s$)$ as opposed to 5 h 17 min $(19,020$ s$)$ suggested by Table I.

## Supplementary Section 2.  ADDITIONAL FIGURES: DETAILED CAPACITIVE COEFFICIENTS IN THE OFF-REGIME

Sensor Gates

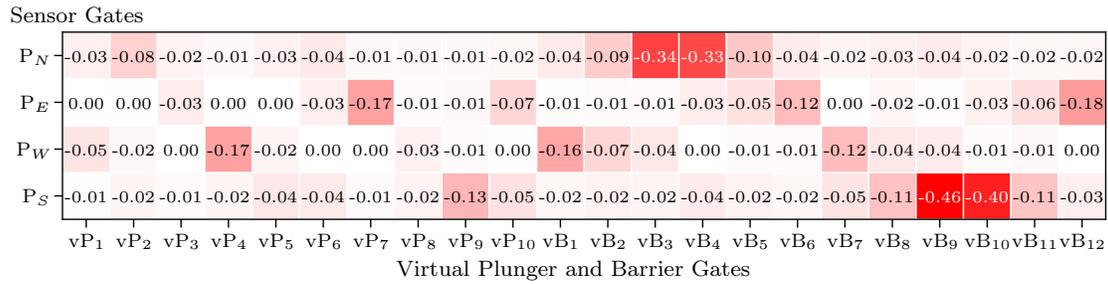

**Supplementary Figure 1.** Charge sensor virtualization. Section of $M_1^{-1}$ characterizing the relative cross-capacitance of each gate to the four charge-sensing plunger gates. The full $M_1^{-1}$ is not shown, as, in practice, the full matrix is a $26 \times 26$ matrix with diagonal entries set to 1, similar to what is shown in the Supplementary Information of Refs. [20,25].

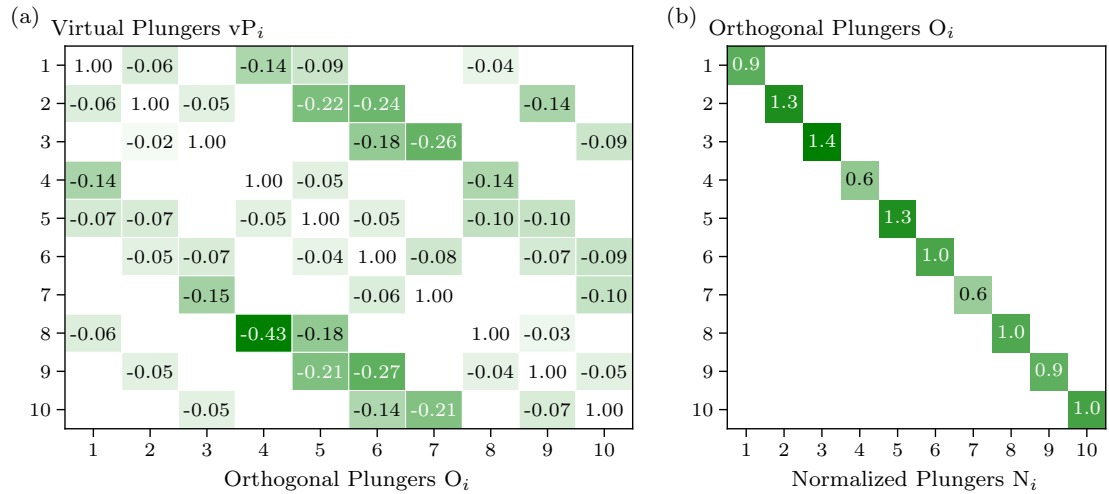

**Supplementary Figure 2.** (a) Plunger gate orthogonalization as depicted in Fig. 2(d). Compensation matrix $M_2^{-1}$ obtained from MAViS for plunger orthogonalization. (b) Plunger gate normalization depicted in Fig. 2(f). Compensation matrix $M_3^{-1}$ obtained from MAViS for plunger normalization.

(a) Normalized Plungers $N_i$

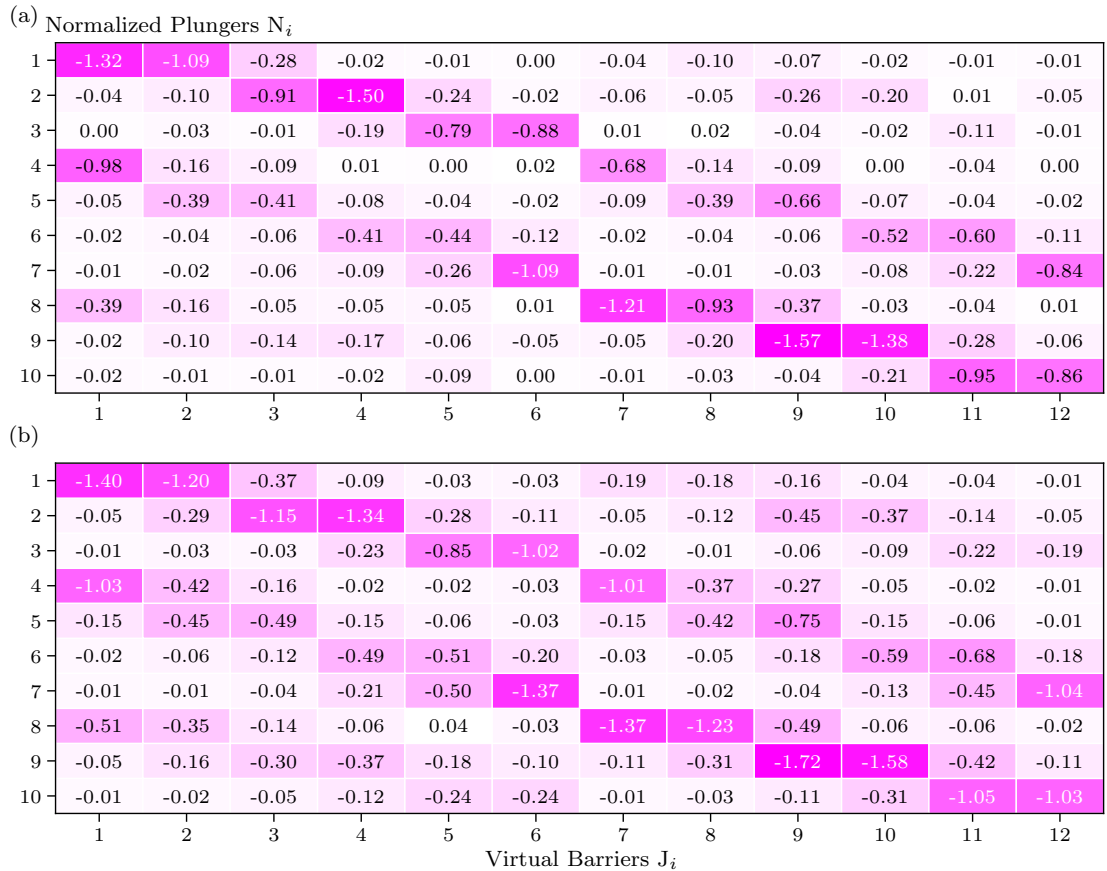

Virtual Barriers $J_i$

**Supplementary Figure 3.** (a) Section of $M_4^{-1}$ showing the barrier-to-plunger virtualization coefficients in the off-coupling regime as depicted in Fig. 3(g). (b) Section of $M_4^{-1}$ including refinements obtained from additional three rounds of measurements. The effect of additional rounds of correction is, on average, $-0.07(8)$ for the first round, $-0.002(36)$ for the second round, and $-0.007(34)$ for the third round. These numbers represent the mean and standard deviation of the full corrections matrices, and their converging trend to a correction matrix with mean of $\approx 0$ suggests two rounds of correction are sufficient for a reliable barrier virtualization. The standard deviation of the last two correction matrices ($\approx 0.03$) can be viewed as an average detection error of our algorithm.

## Supplementary Section 3.  ADDITIONAL FIGURES: DETAILED CAPACITIVE COEFFICIENTS IN THE ON-REGIME

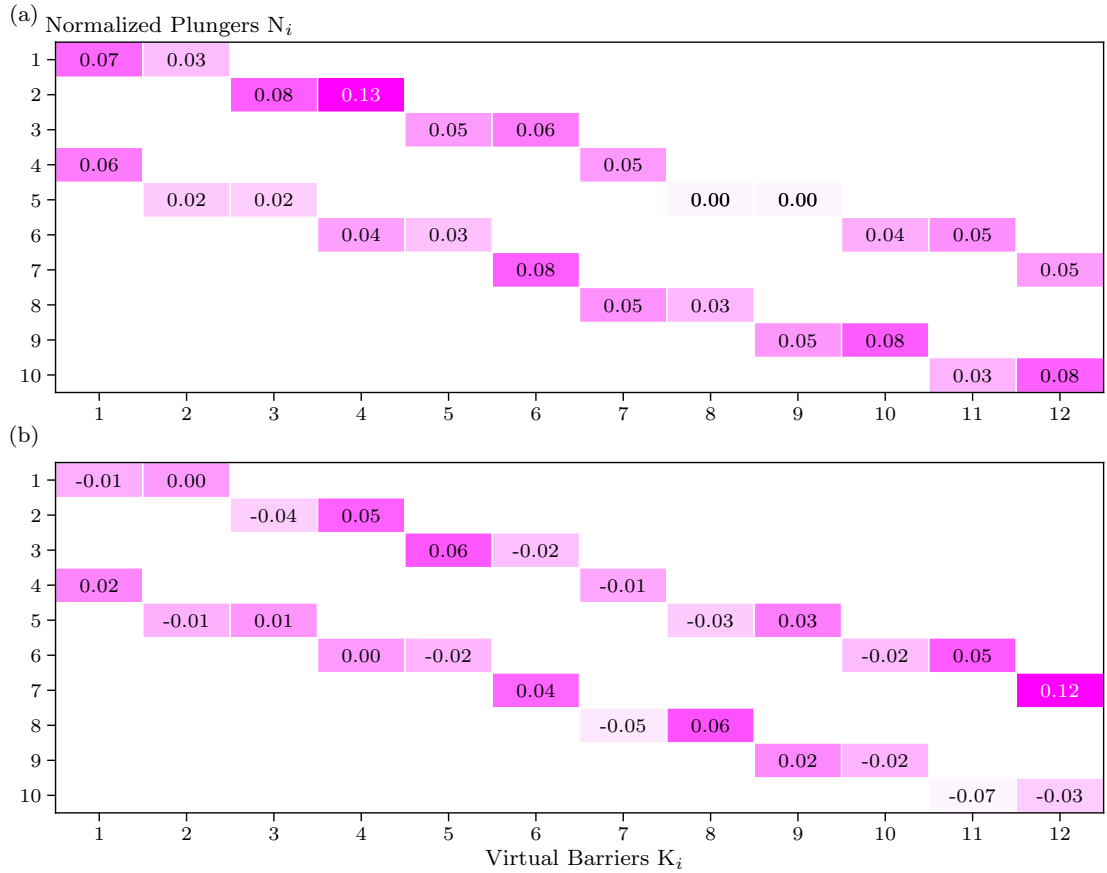

**Supplementary Figure 4**. Barrier virtualization in the on-coupling regime depicted in Fig. 4(f). (a) Matrix showing the square root of the quadratic coefficients ($\sqrt{\alpha}$) in the on-coupling regime. (b) Matrix illustrating the linear coefficients ($\beta$) in the on-coupling regime.




\begin{thebibliography}{71}%
\makeatletter
\providecommand \@ifxundefined [1]{%
 \@ifx{#1\undefined}
}%
\providecommand \@ifnum [1]{%
 \ifnum #1\expandafter \@firstoftwo
 \else \expandafter \@secondoftwo
 \fi
}%
\providecommand \@ifx [1]{%
 \ifx #1\expandafter \@firstoftwo
 \else \expandafter \@secondoftwo
 \fi
}%
\providecommand \natexlab [1]{#1}%
\providecommand \enquote  [1]{``#1''}%
\providecommand \bibnamefont  [1]{#1}%
\providecommand \bibfnamefont [1]{#1}%
\providecommand \citenamefont [1]{#1}%
\providecommand \href@noop [0]{\@secondoftwo}%
\providecommand \href [0]{\begingroup \@sanitize@url \@href}%
\providecommand \@href[1]{\@@startlink{#1}\@@href}%
\providecommand \@@href[1]{\endgroup#1\@@endlink}%
\providecommand \@sanitize@url [0]{\catcode `\\12\catcode `\$12\catcode
  `\&12\catcode `\#12\catcode `\^12\catcode `\_12\catcode `\%12\relax}%
\providecommand \@@startlink[1]{}%
\providecommand \@@endlink[0]{}%
\providecommand \url  [0]{\begingroup\@sanitize@url \@url }%
\providecommand \@url [1]{\endgroup\@href {#1}{\urlprefix }}%
\providecommand \urlprefix  [0]{URL }%
\providecommand \Eprint [0]{\href }%
\providecommand \doibase [0]{https://doi.org/}%
\providecommand \selectlanguage [0]{\@gobble}%
\providecommand \bibinfo  [0]{\@secondoftwo}%
\providecommand \bibfield  [0]{\@secondoftwo}%
\providecommand \translation [1]{[#1]}%
\providecommand \BibitemOpen [0]{}%
\providecommand \bibitemStop [0]{}%
\providecommand \bibitemNoStop [0]{.\EOS\space}%
\providecommand \EOS [0]{\spacefactor3000\relax}%
\providecommand \BibitemShut  [1]{\csname bibitem#1\endcsname}%
\let\auto@bib@innerbib\@empty
\bibitem {Burkard21-SSQ}%
  \BibitemOpen
  \bibfield  {author} {\bibinfo {author} {\bibfnamefont {G.}~\bibnamefont
  {Burkard}}, \bibinfo {author} {\bibfnamefont {T.~D.}\ \bibnamefont {Ladd}},
  \bibinfo {author} {\bibfnamefont {A.}~\bibnamefont {Pan}}, \bibinfo {author}
  {\bibfnamefont {J.~M.}\ \bibnamefont {Nichol}},\ and\ \bibinfo {author}
  {\bibfnamefont {J.~R.}\ \bibnamefont {Petta}},\ }\bibfield  {title} {\bibinfo
  {title} {Semiconductor spin qubits},\ }\href
  {https://doi.org/10.1103/RevModPhys.95.025003} {\bibfield  {journal}
  {\bibinfo  {journal} {Rev. Mod. Phys.}\ }\textbf {\bibinfo {volume} {95}},\
  \bibinfo {pages} {025003} (\bibinfo {year} {2023})}\BibitemShut {NoStop}%
  %
\bibitem {veldhorst2014addressable}%
  \BibitemOpen
  \bibfield  {author} {\bibinfo {author} {\bibfnamefont {M.}~\bibnamefont
  {Veldhorst}}, \bibinfo {author} {\bibfnamefont {J.}~\bibnamefont {Hwang}},
  \bibinfo {author} {\bibfnamefont {C.}~\bibnamefont {Yang}}, \bibinfo {author}
  {\bibfnamefont {A.}~\bibnamefont {Leenstra}}, \bibinfo {author}
  {\bibfnamefont {B.}~\bibnamefont {de~Ronde}}, \bibinfo {author}
  {\bibfnamefont {J.}~\bibnamefont {Dehollain}}, \bibinfo {author}
  {\bibfnamefont {J.}~\bibnamefont {Muhonen}}, \bibinfo {author} {\bibfnamefont
  {F.}~\bibnamefont {Hudson}}, \bibinfo {author} {\bibfnamefont {K.~M.}\
  \bibnamefont {Itoh}}, \bibinfo {author} {\bibfnamefont {A.~t.}\ \bibnamefont
  {Morello}}, \emph {et~al.},\ }\bibfield  {title} {\bibinfo {title} {An
  addressable quantum dot qubit with fault-tolerant control-fidelity},\
  }\href{https://doi.org/10.1038/nnano.2014.216} {\bibfield  {journal} {\bibinfo  {journal} {Nature
  nanotech.}\ }\textbf {\bibinfo {volume} {9}},\ \bibinfo {pages} {981}
  (\bibinfo {year} {2014})}\BibitemShut {NoStop}%
  %
\bibitem {Yoneda2018}%
  \BibitemOpen
  \bibfield  {author} {\bibinfo {author} {\bibfnamefont {J.}~\bibnamefont
  {Yoneda}}, \bibinfo {author} {\bibfnamefont {K.}~\bibnamefont {Takeda}},
  \bibinfo {author} {\bibfnamefont {T.}~\bibnamefont {Otsuka}}, \bibinfo
  {author} {\bibfnamefont {T.}~\bibnamefont {Nakajima}}, \bibinfo {author}
  {\bibfnamefont {M.~R.}\ \bibnamefont {Delbecq}}, \bibinfo {author}
  {\bibfnamefont {G.}~\bibnamefont {Allison}}, \bibinfo {author} {\bibfnamefont
  {T.}~\bibnamefont {Honda}}, \bibinfo {author} {\bibfnamefont
  {T.}~\bibnamefont {Kodera}}, \bibinfo {author} {\bibfnamefont
  {S.}~\bibnamefont {Oda}}, \bibinfo {author} {\bibfnamefont {Y.}~\bibnamefont
  {Hoshi}}, \emph {et~al.},\ }\bibfield
  {title} {\bibinfo {title} {A quantum-dot spin qubit with coherence limited by
  charge noise and fidelity higher than 99.9{\%}},\ }\href
  {https://doi.org/10.1038/s41565-017-0014-x} {\bibfield  {journal} {\bibinfo
  {journal} {Nature Nanotech.}\ }\textbf {\bibinfo {volume} {13}},\
  \bibinfo {pages} {102} (\bibinfo {year} {2018})}\BibitemShut {NoStop}%
  %
\bibitem {Takeda2021}%
  \BibitemOpen
  \bibfield  {author} {\bibinfo {author} {\bibfnamefont {K.}~\bibnamefont
  {Takeda}}, \bibinfo {author} {\bibfnamefont {A.}~\bibnamefont {Noiri}},
  \bibinfo {author} {\bibfnamefont {T.}~\bibnamefont {Nakajima}}, \bibinfo
  {author} {\bibfnamefont {J.}~\bibnamefont {Yoneda}}, \bibinfo {author}
  {\bibfnamefont {T.}~\bibnamefont {Kobayashi}},\ and\ \bibinfo {author}
  {\bibfnamefont {S.}~\bibnamefont {Tarucha}},\ }\bibfield  {title} {\bibinfo
  {title} {Quantum tomography of an entangled three-qubit state in silicon},\
  }\href {https://doi.org/10.1038/s41565-021-00925-0} {\bibfield  {journal}
  {\bibinfo  {journal} {Nature Nanotech.}\ }\textbf {\bibinfo {volume}
  {16}},\ \bibinfo {pages} {965} (\bibinfo {year} {2021})}\BibitemShut
  {NoStop}%
  %
\bibitem {Mills22-TSP}%
  \BibitemOpen
  \bibfield  {author} {\bibinfo {author} {\bibfnamefont {A.~R.}\ \bibnamefont
  {Mills}}, \bibinfo {author} {\bibfnamefont {C.~R.}\ \bibnamefont {Guinn}},
  \bibinfo {author} {\bibfnamefont {M.~J.}\ \bibnamefont {Gullans}}, \bibinfo
  {author} {\bibfnamefont {A.~J.}\ \bibnamefont {Sigillito}}, \bibinfo {author}
  {\bibfnamefont {M.~M.}\ \bibnamefont {Feldman}}, \bibinfo {author}
  {\bibfnamefont {E.}~\bibnamefont {Nielsen}},\ and\ \bibinfo {author}
  {\bibfnamefont {J.~R.}\ \bibnamefont {Petta}},\ }\bibfield  {title} {\bibinfo
  {title} {Two-qubit silicon quantum processor with operation fidelity
  exceeding 99\%},\ }\href {https://doi.org/10.1126/sciadv.abn5130} {\bibfield
  {journal} {\bibinfo  {journal} {Sci. Adv.}\ }\textbf {\bibinfo {volume}
  {8}},\ \bibinfo {pages} {eabn5130} (\bibinfo {year} {2022})}\BibitemShut
  {NoStop}%
  %
\bibitem {Noiri22-FUG}%
  \BibitemOpen
  \bibfield  {author} {\bibinfo {author} {\bibfnamefont {A.}~\bibnamefont
  {Noiri}}, \bibinfo {author} {\bibfnamefont {K.}~\bibnamefont {Takeda}},
  \bibinfo {author} {\bibfnamefont {T.}~\bibnamefont {Nakajima}}, \bibinfo
  {author} {\bibfnamefont {T.}~\bibnamefont {Kobayashi}}, \bibinfo {author}
  {\bibfnamefont {A.}~\bibnamefont {Sammak}}, \bibinfo {author} {\bibfnamefont
  {G.}~\bibnamefont {Scappucci}},\ and\ \bibinfo {author} {\bibfnamefont
  {S.}~\bibnamefont {Tarucha}},\ }\bibfield  {title} {\bibinfo {title} {Fast
  universal quantum gate above the fault-tolerance threshold in silicon},\
  }\href {https://doi.org/10.1038/s41586-021-04182-y} {\bibfield  {journal}
  {\bibinfo  {journal} {Nature}\ }\textbf {\bibinfo {volume} {601}},\ \bibinfo
  {pages} {338} (\bibinfo {year} {2022})}\BibitemShut {NoStop}%
  %
\bibitem {Philips22-UCS}%
  \BibitemOpen
  \bibfield  {author} {\bibinfo {author} {\bibfnamefont {S.~G.~J.}\
  \bibnamefont {Philips}}, \bibinfo {author} {\bibfnamefont {M.~T.}\
  \bibnamefont {Madzik}}, \bibinfo {author} {\bibfnamefont {S.~V.}\
  \bibnamefont {Amitonov}}, \bibinfo {author} {\bibfnamefont {S.~L.}\
  \bibnamefont {de~Snoo}}, \bibinfo {author} {\bibfnamefont {M.}~\bibnamefont
  {Russ}}, \bibinfo {author} {\bibfnamefont {N.}~\bibnamefont {Kalhor}},
  \bibinfo {author} {\bibfnamefont {C.}~\bibnamefont {Volk}}, \bibinfo {author}
  {\bibfnamefont {W.~I.~L.}\ \bibnamefont {Lawrie}}, \bibinfo {author}
  {\bibfnamefont {D.}~\bibnamefont {Brousse}}, \bibinfo {author} {\bibfnamefont
  {L.}~\bibnamefont {Tryputen}}, \emph {et~al.},\ }\bibfield  {title} {\bibinfo {title} {Universal control of a six-qubit
  quantum processor in silicon},\ }\href
  {https://doi.org/10.1038/s41586-022-05117-x} {\bibfield  {journal} {\bibinfo
  {journal} {Nature}\ }\textbf {\bibinfo {volume} {609}},\ \bibinfo {pages}
  {919} (\bibinfo {year} {2022})}\BibitemShut {NoStop}%
  %
\bibitem {Zwerver2022}%
  \BibitemOpen
  \bibfield  {author} {\bibinfo {author} {\bibfnamefont {A.}~\bibnamefont
  {Zwerver}}, \bibinfo {author} {\bibfnamefont {T.}~\bibnamefont
  {Kr{\"a}henmann}}, \bibinfo {author} {\bibfnamefont {T.}~\bibnamefont
  {Watson}}, \bibinfo {author} {\bibfnamefont {L.}~\bibnamefont {Lampert}},
  \bibinfo {author} {\bibfnamefont {H.~C.}\ \bibnamefont {George}}, \bibinfo
  {author} {\bibfnamefont {R.}~\bibnamefont {Pillarisetty}}, \bibinfo {author}
  {\bibfnamefont {S.}~\bibnamefont {Bojarski}}, \bibinfo {author}
  {\bibfnamefont {P.}~\bibnamefont {Amin}}, \bibinfo {author} {\bibfnamefont
  {S.}~\bibnamefont {Amitonov}}, \bibinfo {author} {\bibfnamefont
  {J.}~\bibnamefont {Boter}}, \emph {et~al.},\ }\bibfield  {title} {\bibinfo
  {title} {Qubits made by advanced semiconductor manufacturing},\ }\href
  {https://doi.org/10.1038/s41928-022-00727-9} {\bibfield  {journal} {\bibinfo  {journal} {Nat. Electron.}\ }\textbf
  {\bibinfo {volume} {5}},\ \bibinfo {pages} {184} (\bibinfo {year}
  {2022})}\BibitemShut {NoStop}%
  %
\bibitem {Weinstein2023}%
  \BibitemOpen
  \bibfield  {author} {\bibinfo {author} {\bibfnamefont {A.~J.}\ \bibnamefont
  {Weinstein}}, \bibinfo {author} {\bibfnamefont {M.~D.}\ \bibnamefont {Reed}},
  \bibinfo {author} {\bibfnamefont {A.~M.}\ \bibnamefont {Jones}}, \bibinfo
  {author} {\bibfnamefont {R.~W.}\ \bibnamefont {Andrews}}, \bibinfo {author}
  {\bibfnamefont {D.}~\bibnamefont {Barnes}}, \bibinfo {author} {\bibfnamefont
  {J.~Z.}\ \bibnamefont {Blumoff}}, \bibinfo {author} {\bibfnamefont {L.~E.}\
  \bibnamefont {Euliss}}, \bibinfo {author} {\bibfnamefont {K.}~\bibnamefont
  {Eng}}, \bibinfo {author} {\bibfnamefont {B.~H.}\ \bibnamefont {Fong}},
  \bibinfo {author} {\bibfnamefont {S.~D.}\ \bibnamefont {Ha}}, \bibinfo
  {author} {\bibfnamefont {D.~R.}\ \bibnamefont {Hulbert}}, \emph {et~al.},\ }\bibfield  {title}
  {\bibinfo {title} {Universal logic with encoded spin qubits in silicon},\
  }\href {https://doi.org/10.1038/s41586-023-05777-3} {\bibfield  {journal}
  {\bibinfo  {journal} {Nature}\ }\textbf {\bibinfo {volume} {615}},\ \bibinfo
  {pages} {817} (\bibinfo {year} {2023})}\BibitemShut {NoStop}%
  %
\bibitem {Desmet2024}%
  \BibitemOpen
  \bibfield  {author} {\bibinfo {author} {\bibfnamefont {M.~D.}\ \bibnamefont
  {Smet}}, \bibinfo {author} {\bibfnamefont {Y.}~\bibnamefont {Matsumoto}},
  \bibinfo {author} {\bibfnamefont {A.-M.~J.}\ \bibnamefont {Zwerver}},
  \bibinfo {author} {\bibfnamefont {L.}~\bibnamefont {Tryputen}}, \bibinfo
  {author} {\bibfnamefont {S.~L.}\ \bibnamefont {de~Snoo}}, \bibinfo {author}
  {\bibfnamefont {S.~V.}\ \bibnamefont {Amitonov}}, \bibinfo {author}
  {\bibfnamefont {A.}~\bibnamefont {Sammak}}, \bibinfo {author} {\bibfnamefont
  {N.}~\bibnamefont {Samkharadze}}, \bibinfo {author} {\bibnamefont {Ö.
  Gül}}, \bibinfo {author} {\bibfnamefont {R.~N.~M.}\ \bibnamefont
  {Wasserman}}, \emph {et~al.},\ }{\bibinfo {title}
  {High-fidelity single-spin shuttling in silicon}},\
  \Eprint {https://arxiv.org/abs/2406.07267} {arXiv:2406.07267} (\bibinfo {year} {2024})\BibitemShut {NoStop}%
  %
\bibitem {Huang2024}%
  \BibitemOpen
  \bibfield  {author} {\bibinfo {author} {\bibfnamefont {J.~Y.}\ \bibnamefont
  {Huang}}, \bibinfo {author} {\bibfnamefont {R.~Y.}\ \bibnamefont {Su}},
  \bibinfo {author} {\bibfnamefont {W.~H.}\ \bibnamefont {Lim}}, \bibinfo
  {author} {\bibfnamefont {M.}~\bibnamefont {Feng}}, \bibinfo {author}
  {\bibfnamefont {B.}~\bibnamefont {van Straaten}}, \bibinfo {author}
  {\bibfnamefont {B.}~\bibnamefont {Severin}}, \bibinfo {author} {\bibfnamefont
  {W.}~\bibnamefont {Gilbert}}, \bibinfo {author} {\bibfnamefont
  {N.}~\bibnamefont {Dumoulin~Stuyck}}, \bibinfo {author} {\bibfnamefont
  {T.}~\bibnamefont {Tanttu}}, \bibinfo {author} {\bibfnamefont
  {S.}~\bibnamefont {Serrano}}, \emph {et~al.},\ }\bibfield  {title} {\bibinfo {title} {High-fidelity
  spin qubit operation and algorithmic initialization above 1K},\ }\href
  {https://doi.org/10.1038/s41586-024-07160-2} {\bibfield  {journal} {\bibinfo
  {journal} {Nature}\ }\textbf {\bibinfo {volume} {627}},\ \bibinfo {pages}
  {772} (\bibinfo {year} {2024})}\BibitemShut {NoStop}%
  %
\bibitem {Neyens2024}%
  \BibitemOpen
  \bibfield  {author} {\bibinfo {author} {\bibfnamefont {S.}~\bibnamefont
  {Neyens}}, \bibinfo {author} {\bibfnamefont {O.~K.}\ \bibnamefont {Zietz}},
  \bibinfo {author} {\bibfnamefont {T.~F.}\ \bibnamefont {Watson}}, \bibinfo
  {author} {\bibfnamefont {F.}~\bibnamefont {Luthi}}, \bibinfo {author}
  {\bibfnamefont {A.}~\bibnamefont {Nethwewala}}, \bibinfo {author}
  {\bibfnamefont {H.~C.}\ \bibnamefont {George}}, \bibinfo {author}
  {\bibfnamefont {E.}~\bibnamefont {Henry}}, \bibinfo {author} {\bibfnamefont
  {M.}~\bibnamefont {Islam}}, \bibinfo {author} {\bibfnamefont {A.~J.}\
  \bibnamefont {Wagner}}, \bibinfo {author} {\bibfnamefont {F.}~\bibnamefont
  {Borjans}}, \emph {et~al.},\ }\bibfield  {title} {\bibinfo {title} {Probing single electrons across 300-mm
  spin qubit wafers},\ }\href {https://doi.org/10.1038/s41586-024-07275-6}
  {\bibfield  {journal} {\bibinfo  {journal} {Nature}\ }\textbf {\bibinfo
  {volume} {629}},\ \bibinfo {pages} {80} (\bibinfo {year} {2024})}\BibitemShut
  {NoStop}%
  %
\bibitem {Scappucci2021}%
  \BibitemOpen
  \bibfield  {author} {\bibinfo {author} {\bibfnamefont {G.}~\bibnamefont
  {Scappucci}}, \bibinfo {author} {\bibfnamefont {C.}~\bibnamefont {Kloeffel}},
  \bibinfo {author} {\bibfnamefont {F.~A.}\ \bibnamefont {Zwanenburg}},
  \bibinfo {author} {\bibfnamefont {D.}~\bibnamefont {Loss}}, \bibinfo {author}
  {\bibfnamefont {M.}~\bibnamefont {Myronov}}, \bibinfo {author} {\bibfnamefont
  {J.-J.}\ \bibnamefont {Zhang}}, \bibinfo {author} {\bibfnamefont
  {S.}~\bibnamefont {De~Franceschi}}, \bibinfo {author} {\bibfnamefont
  {G.}~\bibnamefont {Katsaros}},\ and\ \bibinfo {author} {\bibfnamefont
  {M.}~\bibnamefont {Veldhorst}},\ }\bibfield  {title} {\bibinfo {title} {The
  germanium quantum information route},\ }\href
  {https://doi.org/10.1038/s41578-020-00262-z} {\bibfield  {journal} {\bibinfo
  {journal} {Nat. Rev. Mater.}\ }\textbf {\bibinfo {volume} {6}},\ \bibinfo
  {pages} {926} (\bibinfo {year} {2021})}\BibitemShut {NoStop}%
  %
\bibitem{Watzinger2018}%
  \BibitemOpen
  \bibfield  {author} {\bibinfo {author} {\bibfnamefont {H.}~\bibnamefont
  {Watzinger}}, \bibinfo {author} {\bibfnamefont {J.}~\bibnamefont
  {Kuku{\v{c}}ka}}, \bibinfo {author} {\bibfnamefont {L.}~\bibnamefont
  {Vuku{\v{s}}i{\'{c}}}}, \bibinfo {author} {\bibfnamefont {F.}~\bibnamefont
  {Gao}}, \bibinfo {author} {\bibfnamefont {T.}~\bibnamefont {Wang}}, \bibinfo
  {author} {\bibfnamefont {F.}~\bibnamefont {Sch{\"a}ffler}}, \bibinfo {author}
  {\bibfnamefont {J.-J.}\ \bibnamefont {Zhang}},\ and\ \bibinfo {author}
  {\bibfnamefont {G.}~\bibnamefont {Katsaros}},\ }\bibfield  {title} {\bibinfo
  {title} {A germanium hole spin qubit},\ }\href
  {https://doi.org/10.1038/s41467-018-06418-4} {\bibfield  {journal} {\bibinfo
  {journal} {Nat. Commun.}\ }\textbf {\bibinfo {volume} {9}},\
  \bibinfo {pages} {3902} (\bibinfo {year} {2018})}\BibitemShut {NoStop}%
  %
\bibitem {Lodari2021}%
  \BibitemOpen
  \bibfield  {author} {\bibinfo {author} {\bibfnamefont {M.}~\bibnamefont
  {Lodari}}, \bibinfo {author} {\bibfnamefont {N.~W.}\ \bibnamefont
  {Hendrickx}}, \bibinfo {author} {\bibfnamefont {W.~I.~L.}\ \bibnamefont
  {Lawrie}}, \bibinfo {author} {\bibfnamefont {T.-K.}\ \bibnamefont {Hsiao}},
  \bibinfo {author} {\bibfnamefont {L.~M.~K.}\ \bibnamefont {Vandersypen}},
  \bibinfo {author} {\bibfnamefont {A.}~\bibnamefont {Sammak}}, \bibinfo
  {author} {\bibfnamefont {M.}~\bibnamefont {Veldhorst}},\ and\ \bibinfo
  {author} {\bibfnamefont {G.}~\bibnamefont {Scappucci}},\ }\bibfield  {title}
  {\bibinfo {title} {Low percolation density and charge noise with holes in
  germanium},\ }\href {https://doi.org/10.1088/2633-4356/abcd82} {\bibfield
  {journal} {\bibinfo  {journal} {Mater. Quantum Technol.}\ }\textbf
  {\bibinfo {volume} {1}},\ \bibinfo {pages} {011002} (\bibinfo {year}
  {2021})}\BibitemShut {NoStop}%
  %
\bibitem {Sammak2019}%
  \BibitemOpen
  \bibfield  {author} {\bibinfo {author} {\bibfnamefont {A.}~\bibnamefont
  {Sammak}}, \bibinfo {author} {\bibfnamefont {D.}~\bibnamefont {Sabbagh}},
  \bibinfo {author} {\bibfnamefont {N.~W.}\ \bibnamefont {Hendrickx}}, \bibinfo
  {author} {\bibfnamefont {M.}~\bibnamefont {Lodari}}, \bibinfo {author}
  {\bibfnamefont {B.}~\bibnamefont {Paquelet~Wuetz}}, \bibinfo {author}
  {\bibfnamefont {A.}~\bibnamefont {Tosato}}, \bibinfo {author} {\bibfnamefont
  {L.}~\bibnamefont {Yeoh}}, \bibinfo {author} {\bibfnamefont {M.}~\bibnamefont
  {Bollani}}, \bibinfo {author} {\bibfnamefont {M.}~\bibnamefont {Virgilio}},
  \bibinfo {author} {\bibfnamefont {M.~A.}\ \bibnamefont {Schubert}}, \emph {et~al.},\ }\bibfield  {title} {\bibinfo
  {title} {Shallow and {Undoped} {Germanium} {Quantum} {Wells}: {A}
  {Playground} for {Spin} and {Hybrid} {Quantum} {Technology}},\ }\href
  {https://doi.org/10.1002/adfm.201807613} {\bibfield  {journal} {\bibinfo
  {journal} {Adv. Funct. Mater.}\ }\textbf {\bibinfo {volume}
  {29}},\ \bibinfo {pages} {1807613} (\bibinfo {year} {2019})}\BibitemShut
  {NoStop}%
  %
\bibitem {Hendrickx2018}%
  \BibitemOpen
  \bibfield  {author} {\bibinfo {author} {\bibfnamefont {N.~W.}\ \bibnamefont
  {Hendrickx}}, \bibinfo {author} {\bibfnamefont {D.~P.}\ \bibnamefont
  {Franke}}, \bibinfo {author} {\bibfnamefont {A.}~\bibnamefont {Sammak}},
  \bibinfo {author} {\bibfnamefont {M.}~\bibnamefont {Kouwenhoven}}, \bibinfo
  {author} {\bibfnamefont {D.}~\bibnamefont {Sabbagh}}, \bibinfo {author}
  {\bibfnamefont {L.}~\bibnamefont {Yeoh}}, \bibinfo {author} {\bibfnamefont
  {R.}~\bibnamefont {Li}}, \bibinfo {author} {\bibfnamefont {M.~L.~V.}\
  \bibnamefont {Tagliaferri}}, \bibinfo {author} {\bibfnamefont
  {M.}~\bibnamefont {Virgilio}}, \bibinfo {author} {\bibfnamefont
  {G.}~\bibnamefont {Capellini}}, \emph {et~al.},\ }\bibfield  {title} {\bibinfo {title}
  {Gate-controlled quantum dots and superconductivity in planar germanium},\
  }\href {https://doi.org/10.1038/s41467-018-05299-x} {\bibfield  {journal}
  {\bibinfo  {journal} {Nat. Commun.}\ }\textbf {\bibinfo {volume}
  {9}},\ \bibinfo {pages} {2835} (\bibinfo {year} {2018})}\BibitemShut
  {NoStop}%
  %
\bibitem {Stehouwer2023}%
  \BibitemOpen
  \bibfield  {author} {\bibinfo {author} {\bibfnamefont {L.~E.~A.}\
  \bibnamefont {Stehouwer}}, \bibinfo {author} {\bibfnamefont {A.}~\bibnamefont
  {Tosato}}, \bibinfo {author} {\bibfnamefont {D.}~\bibnamefont
  {Degli~Esposti}}, \bibinfo {author} {\bibfnamefont {D.}~\bibnamefont
  {Costa}}, \bibinfo {author} {\bibfnamefont {M.}~\bibnamefont {Veldhorst}},
  \bibinfo {author} {\bibfnamefont {A.}~\bibnamefont {Sammak}},\ and\ \bibinfo
  {author} {\bibfnamefont {G.}~\bibnamefont {Scappucci}},\ }\bibfield  {title}
  {\bibinfo {title} {{Germanium wafers for strained quantum wells with low
  disorder}},\ }\href {https://doi.org/10.1063/5.0158262} {\bibfield  {journal}
  {\bibinfo  {journal} {Appl. Phys. Lett.}\ }\textbf {\bibinfo {volume}
  {123}},\ \bibinfo {pages} {092101} (\bibinfo {year} {2023})}  \BibitemShut {NoStop}%
  %
\bibitem {vanRiggelen2021}%
  \BibitemOpen
  \bibfield  {author} {\bibinfo {author} {\bibfnamefont {F.}~\bibnamefont {van
  Riggelen}}, \bibinfo {author} {\bibfnamefont {N.~W.}\ \bibnamefont
  {Hendrickx}}, \bibinfo {author} {\bibfnamefont {W.~I.~L.}\ \bibnamefont
  {Lawrie}}, \bibinfo {author} {\bibfnamefont {M.}~\bibnamefont {Russ}},
  \bibinfo {author} {\bibfnamefont {A.}~\bibnamefont {Sammak}}, \bibinfo
  {author} {\bibfnamefont {G.}~\bibnamefont {Scappucci}},\ and\ \bibinfo
  {author} {\bibfnamefont {M.}~\bibnamefont {Veldhorst}},\ }\bibfield  {title}
  {\bibinfo {title} {{A two-dimensional array of single-hole quantum dots}},\
  }\href {https://doi.org/10.1063/5.0037330} {\bibfield  {journal} {\bibinfo
  {journal} {Appl. Phys. Lett.}\ }\textbf {\bibinfo {volume} {118}},\
  \bibinfo {pages} {044002} (\bibinfo {year} {2021})}
  \BibitemShut {NoStop}%
  %
\bibitem {Borsoi22-QCA}%
  \BibitemOpen
  \bibfield  {author} {\bibinfo {author} {\bibfnamefont {F.}~\bibnamefont
  {Borsoi}}, \bibinfo {author} {\bibfnamefont {N.~W.}\ \bibnamefont
  {Hendrickx}}, \bibinfo {author} {\bibfnamefont {V.}~\bibnamefont {John}},
  \bibinfo {author} {\bibfnamefont {M.}~\bibnamefont {Meyer}}, \bibinfo
  {author} {\bibfnamefont {S.}~\bibnamefont {Motz}}, \bibinfo {author}
  {\bibfnamefont {F.}~\bibnamefont {van Riggelen}}, \bibinfo {author}
  {\bibfnamefont {A.}~\bibnamefont {Sammak}}, \bibinfo {author} {\bibfnamefont
  {S.~L.}\ \bibnamefont {de~Snoo}}, \bibinfo {author} {\bibfnamefont
  {G.}~\bibnamefont {Scappucci}},\ and\ \bibinfo {author} {\bibfnamefont
  {M.}~\bibnamefont {Veldhorst}},\ }\bibfield  {title} {\bibinfo {title}
  {Shared control of a 16 semiconductor quantum dot crossbar array},\ }\href
  {https://doi.org/10.1038/s41565-023-01491-3} {\bibfield  {journal} {\bibinfo
  {journal} {Nat. Nanotechnol.}\ } (\bibinfo {year} {2023})}\BibitemShut
  {NoStop}%
  %
\bibitem {Hsiao2024}%
  \BibitemOpen
  \bibfield  {author} {\bibinfo {author} {\bibfnamefont {T.-K.}\ \bibnamefont
  {Hsiao}}, \bibinfo {author} {\bibfnamefont {P.}~\bibnamefont {Cova
  Fari\~na}}, \bibinfo {author} {\bibfnamefont {S.~D.}\ \bibnamefont
  {Oosterhout}}, \bibinfo {author} {\bibfnamefont {D.}~\bibnamefont {Jirovec}},
  \bibinfo {author} {\bibfnamefont {X.}~\bibnamefont {Zhang}}, \bibinfo
  {author} {\bibfnamefont {C.~J.}\ \bibnamefont {van Diepen}}, \bibinfo
  {author} {\bibfnamefont {W.~I.~L.}\ \bibnamefont {Lawrie}}, \bibinfo {author}
  {\bibfnamefont {C.-A.}\ \bibnamefont {Wang}}, \bibinfo {author}
  {\bibfnamefont {A.}~\bibnamefont {Sammak}}, \bibinfo {author} {\bibfnamefont
  {G.}~\bibnamefont {Scappucci}}, \emph {et~al.},\ }\bibfield  {title} {\bibinfo
  {title} {Exciton transport in a germanium quantum dot ladder},\ }\href
  {https://doi.org/10.1103/PhysRevX.14.011048} {\bibfield  {journal} {\bibinfo
  {journal} {Phys. Rev. X}\ }\textbf {\bibinfo {volume} {14}},\ \bibinfo
  {pages} {011048} (\bibinfo {year} {2024})}\BibitemShut {NoStop}%
  %
\bibitem {Hendrickx_2021}%
  \BibitemOpen
  \bibfield  {author} {\bibinfo {author} {\bibfnamefont {N.~W.}\ \bibnamefont
  {Hendrickx}}, \bibinfo {author} {\bibfnamefont {W.~I.~L.}\ \bibnamefont
  {Lawrie}}, \bibinfo {author} {\bibfnamefont {M.}~\bibnamefont {Russ}},
  \bibinfo {author} {\bibfnamefont {F.}~\bibnamefont {van Riggelen}}, \bibinfo
  {author} {\bibfnamefont {S.~L.}\ \bibnamefont {de~Snoo}}, \bibinfo {author}
  {\bibfnamefont {R.~N.}\ \bibnamefont {Schouten}}, \bibinfo {author}
  {\bibfnamefont {A.}~\bibnamefont {Sammak}}, \bibinfo {author} {\bibfnamefont
  {G.}~\bibnamefont {Scappucci}},\ and\ \bibinfo {author} {\bibfnamefont
  {M.}~\bibnamefont {Veldhorst}},\ }\bibfield  {title} {\bibinfo {title} {A
  four-qubit germanium quantum processor},\ }\href
  {https://doi.org/10.1038/s41586-021-03332-6} {\bibfield  {journal} {\bibinfo
  {journal} {Nature}\ }\textbf {\bibinfo {volume} {591}},\ \bibinfo {pages}
  {580–585} (\bibinfo {year} {2021})}\BibitemShut {NoStop}%
  %
\bibitem {Zhang2023}%
  \BibitemOpen
  \bibfield  {author} {\bibinfo {author} {\bibfnamefont {X.}~\bibnamefont
  {Zhang}}, \bibinfo {author} {\bibfnamefont {E.}~\bibnamefont {Morozova}},
  \bibinfo {author} {\bibfnamefont {M.}~\bibnamefont {Rimbach-Russ}}, \bibinfo
  {author} {\bibfnamefont {D.}~\bibnamefont {Jirovec}}, \bibinfo {author}
  {\bibfnamefont {T.-K.}\ \bibnamefont {Hsiao}}, \bibinfo {author}
  {\bibfnamefont {P.~C.}\ \bibnamefont {Fari{\~n}a}}, \bibinfo {author}
  {\bibfnamefont {C.-A.}\ \bibnamefont {Wang}}, \bibinfo {author}
  {\bibfnamefont {S.~D.}\ \bibnamefont {Oosterhout}}, \bibinfo {author}
  {\bibfnamefont {A.}~\bibnamefont {Sammak}}, \bibinfo {author} {\bibfnamefont
  {G.}~\bibnamefont {Scappucci}}, \emph {et~al.},\ }\bibfield  {title}
  {\bibinfo {title} {Universal control of four singlet-triplet qubits}},\
  \Eprint {https://arxiv.org/abs/arXiv:2312.16101} {arXiv:2312.16101} (\bibinfo {year} {2023})\BibitemShut {NoStop}%
  %
\bibitem {Wang2024}%
  \BibitemOpen
  \bibfield  {author} {\bibinfo {author} {\bibfnamefont {C.-A.}\ \bibnamefont
  {Wang}}, \bibinfo {author} {\bibfnamefont {V.}~\bibnamefont {John}}, \bibinfo
  {author} {\bibfnamefont {H.}~\bibnamefont {Tidjani}}, \bibinfo {author}
  {\bibfnamefont {C.~X.}\ \bibnamefont {Yu}}, \bibinfo {author} {\bibfnamefont
  {A.~S.}\ \bibnamefont {Ivlev}}, \bibinfo {author} {\bibfnamefont
  {C.}~\bibnamefont {Déprez}}, \bibinfo {author} {\bibfnamefont
  {F.}~\bibnamefont {van Riggelen-Doelman}}, \bibinfo {author} {\bibfnamefont
  {B.~D.}\ \bibnamefont {Woods}}, \bibinfo {author} {\bibfnamefont {N.~W.}\
  \bibnamefont {Hendrickx}}, \bibinfo {author} {\bibfnamefont {W.~I.~L.}\
  \bibnamefont {Lawrie}}, \emph {et~al.},\ }\bibfield  {title} {\bibinfo {title} {Operating semiconductor
  quantum processors with hopping spins},\ }\href
  {https://doi.org/10.1126/science.ado5915} {\bibfield  {journal} {\bibinfo
  {journal} {Science}\ }\textbf {\bibinfo {volume} {385}},\ \bibinfo {pages}
  {447} (\bibinfo {year} {2024})} \BibitemShut
  {NoStop}%
  %
\bibitem {Ha22-FDQ}%
  \BibitemOpen
  \bibfield  {author} {\bibinfo {author} {\bibfnamefont {W.}~\bibnamefont
  {Ha}}, \bibinfo {author} {\bibfnamefont {S.~D.}\ \bibnamefont {Ha}}, \bibinfo
  {author} {\bibfnamefont {M.~D.}\ \bibnamefont {Choi}}, \bibinfo {author}
  {\bibfnamefont {Y.}~\bibnamefont {Tang}}, \bibinfo {author} {\bibfnamefont
  {A.~E.}\ \bibnamefont {Schmitz}}, \bibinfo {author} {\bibfnamefont {M.~P.}\
  \bibnamefont {Levendorf}}, \bibinfo {author} {\bibfnamefont {K.}~\bibnamefont
  {Lee}}, \bibinfo {author} {\bibfnamefont {J.~M.}\ \bibnamefont {Chappell}},
  \bibinfo {author} {\bibfnamefont {T.~S.}\ \bibnamefont {Adams}}, \bibinfo
  {author} {\bibfnamefont {D.~R.}\ \bibnamefont {Hulbert}}, \emph {et~al.},\ 
  }\bibfield  {title} {\bibinfo {title} {A flexible design platform for si/sige
  exchange-only qubits with low disorder},\ }\href
  {https://doi.org/10.1021/acs.nanolett.1c03026} {\bibfield  {journal}
  {\bibinfo  {journal} {Nano Lett.}\ }\textbf {\bibinfo {volume} {22}},\
  \bibinfo {pages} {1443} (\bibinfo {year} {2022})}\BibitemShut {NoStop}%
  %
\bibitem {George2024}%
  \BibitemOpen
  \bibfield  {author} {\bibinfo {author} {\bibfnamefont {H.~C.}\ \bibnamefont
  {George}}, \bibinfo {author} {\bibfnamefont {M.~T.}\ \bibnamefont {Mądzik}},
  \bibinfo {author} {\bibfnamefont {E.~M.}\ \bibnamefont {Henry}}, \bibinfo
  {author} {\bibfnamefont {A.~J.}\ \bibnamefont {Wagner}}, \bibinfo {author}
  {\bibfnamefont {M.~M.}\ \bibnamefont {Islam}}, \bibinfo {author}
  {\bibfnamefont {F.}~\bibnamefont {Borjans}}, \bibinfo {author} {\bibfnamefont
  {E.~J.}\ \bibnamefont {Connors}}, \bibinfo {author} {\bibfnamefont
  {J.}~\bibnamefont {Corrigan}}, \bibinfo {author} {\bibfnamefont
  {M.}~\bibnamefont {Curry}}, \bibinfo {author} {\bibfnamefont {M.~K.}\
  \bibnamefont {Harper}}, \emph {et~al.},\ }{\bibinfo {title}
  {12-spin-qubit arrays fabricated on a 300 mm semiconductor manufacturing
  line}},\ \Eprint {https://arxiv.org/abs/2410.16583}
  {arXiv:2410.16583} (\bibinfo {year} {2024})\BibitemShut {NoStop}%
  %
\bibitem {Durrer2020}%
  \BibitemOpen
  \bibfield  {author} {\bibinfo {author} {\bibfnamefont {R.}~\bibnamefont
  {Durrer}}, \bibinfo {author} {\bibfnamefont {B.}~\bibnamefont {Kratochwil}},
  \bibinfo {author} {\bibfnamefont {J.}~\bibnamefont {Koski}}, \bibinfo
  {author} {\bibfnamefont {A.}~\bibnamefont {Landig}}, \bibinfo {author}
  {\bibfnamefont {C.}~\bibnamefont {Reichl}}, \bibinfo {author} {\bibfnamefont
  {W.}~\bibnamefont {Wegscheider}}, \bibinfo {author} {\bibfnamefont
  {T.}~\bibnamefont {Ihn}},\ and\ \bibinfo {author} {\bibfnamefont
  {E.}~\bibnamefont {Greplova}},\ }\bibfield  {title} {\bibinfo {title}
  {Automated tuning of double quantum dots into specific charge states using
  neural networks},\ }\href {https://doi.org/10.1103/PhysRevApplied.13.054019}
  {\bibfield  {journal} {\bibinfo  {journal} {Phys. Rev. Appl.}\ }\textbf
  {\bibinfo {volume} {13}},\ \bibinfo {pages} {054019} (\bibinfo {year}
  {2020})}\BibitemShut {NoStop}%
  %
\bibitem {Zwolak20-AQD}%
  \BibitemOpen
  \bibfield  {author} {\bibinfo {author} {\bibfnamefont {J.~P.}\ \bibnamefont
  {Zwolak}}, \bibinfo {author} {\bibfnamefont {T.}~\bibnamefont {McJunkin}},
  \bibinfo {author} {\bibfnamefont {S.~S.}\ \bibnamefont {Kalantre}}, \bibinfo
  {author} {\bibfnamefont {J.}~\bibnamefont {Dodson}}, \bibinfo {author}
  {\bibfnamefont {E.~R.}\ \bibnamefont {MacQuarrie}}, \bibinfo {author}
  {\bibfnamefont {D.}~\bibnamefont {Savage}}, \bibinfo {author} {\bibfnamefont
  {M.}~\bibnamefont {Lagally}}, \bibinfo {author} {\bibfnamefont
  {S.}~\bibnamefont {Coppersmith}}, \bibinfo {author} {\bibfnamefont {M.~A.}\
  \bibnamefont {Eriksson}},\ and\ \bibinfo {author} {\bibfnamefont {J.~M.}\
  \bibnamefont {Taylor}},\ }\bibfield  {title} {\bibinfo {title} {Autotuning of
  double-dot devices in situ with machine learning},\ }\href
  {https://doi.org/10.1103/PhysRevApplied.13.034075} {\bibfield  {journal}
  {\bibinfo  {journal} {Phys. Rev. Appl.}\ }\textbf {\bibinfo {volume} {13}},\
  \bibinfo {pages} {034075} (\bibinfo {year} {2020})}\BibitemShut {NoStop}%
  %
\bibitem {Ziegler22-TAR}%
  \BibitemOpen
  \bibfield  {author} {\bibinfo {author} {\bibfnamefont {J.}~\bibnamefont
  {Ziegler}}, \bibinfo {author} {\bibfnamefont {F.}~\bibnamefont {Luthi}},
  \bibinfo {author} {\bibfnamefont {M.}~\bibnamefont {Ramsey}}, \bibinfo
  {author} {\bibfnamefont {F.}~\bibnamefont {Borjans}}, \bibinfo {author}
  {\bibfnamefont {G.}~\bibnamefont {Zheng}},\ and\ \bibinfo {author}
  {\bibfnamefont {J.~P.}\ \bibnamefont {Zwolak}},\ }\bibfield  {title}
  {\bibinfo {title} {Tuning arrays with rays: Physics-informed tuning of
  quantum dot charge states},\ }\href
  {https://doi.org/10.1103/PhysRevApplied.20.034067} {\bibfield  {journal}
  {\bibinfo  {journal} {Phys. Rev. Appl.}\ }\textbf {\bibinfo {volume} {20}},\
  \bibinfo {pages} {034067} (\bibinfo {year} {2023}{\natexlab{a}})}\BibitemShut
  {NoStop}%
  %
\bibitem {Zubchenko24-ABQ}%
  \BibitemOpen
  \bibfield  {author} {\bibinfo {author} {\bibfnamefont {A.}~\bibnamefont
  {Zubchenko}}, \bibinfo {author} {\bibfnamefont {D.}~\bibnamefont
  {Middlebrooks}}, \bibinfo {author} {\bibfnamefont {T.}~\bibnamefont
  {Rasmussen}}, \bibinfo {author} {\bibfnamefont {L.}~\bibnamefont {Lausen}},
  \bibinfo {author} {\bibfnamefont {F.}~\bibnamefont {Kuemmeth}}, \bibinfo
  {author} {\bibfnamefont {A.}~\bibnamefont {Chatterjee}},\ and\ \bibinfo
  {author} {\bibfnamefont {J.~P.}\ \bibnamefont {Zwolak}},\ }\bibfield  {title}
  {\bibinfo {title} {Autonomous bootstrapping of quantum dot devices},\ }\href
  {https://doi.org/10.48550/arXiv.2407.20061} {\bibfield  {journal} {\bibinfo
  {journal} {arXiv:2407.20061}\ }(\bibinfo {year} {2024})}\BibitemShut
  {NoStop}%
  %
\bibitem {Kalantre17-MLD}%
  \BibitemOpen
  \bibfield  {author} {\bibinfo {author} {\bibfnamefont {S.~S.}\ \bibnamefont
  {Kalantre}}, \bibinfo {author} {\bibfnamefont {J.~P.}\ \bibnamefont
  {Zwolak}}, \bibinfo {author} {\bibfnamefont {S.}~\bibnamefont {Ragole}},
  \bibinfo {author} {\bibfnamefont {X.}~\bibnamefont {Wu}}, \bibinfo {author}
  {\bibfnamefont {N.~M.}\ \bibnamefont {Zimmerman}}, \bibinfo {author}
  {\bibfnamefont {M.~D.}\ \bibnamefont {Stewart}},\ and\ \bibinfo {author}
  {\bibfnamefont {J.~M.}\ \bibnamefont {Taylor}},\ }\bibfield  {title}
  {\bibinfo {title} {Machine learning techniques for state recognition and
  auto-tuning in quantum dots},\ }\href
  {https://doi.org/10.1038/s41534-018-0118-7} {\bibfield  {journal} {\bibinfo
  {journal} {npj Quantum Inf.}\ }\textbf {\bibinfo {volume} {5}},\ \bibinfo
  {pages} {1} (\bibinfo {year} {2019})}\BibitemShut {NoStop}%
  %
\bibitem {Zwolak21-RBI}%
  \BibitemOpen
  \bibfield  {author} {\bibinfo {author} {\bibfnamefont {J.~P.}\ \bibnamefont
  {Zwolak}}, \bibinfo {author} {\bibfnamefont {T.}~\bibnamefont {McJunkin}},
  \bibinfo {author} {\bibfnamefont {S.~S.}\ \bibnamefont {Kalantre}}, \bibinfo
  {author} {\bibfnamefont {S.~F.}\ \bibnamefont {Neyens}}, \bibinfo {author}
  {\bibfnamefont {E.~R.}\ \bibnamefont {MacQuarrie}}, \bibinfo {author}
  {\bibfnamefont {M.~A.}\ \bibnamefont {Eriksson}},\ and\ \bibinfo {author}
  {\bibfnamefont {J.~M.}\ \bibnamefont {Taylor}},\ }\bibfield  {title}
  {\bibinfo {title} {Ray-based framework for state identification in quantum
  dot devices},\ }\href {https://doi.org/10.1103/PRXQuantum.2.020335}
  {\bibfield  {journal} {\bibinfo  {journal} {PRX Quantum}\ }\textbf {\bibinfo
  {volume} {2}},\ \bibinfo {pages} {020335} (\bibinfo {year}
  {2021})}\BibitemShut {NoStop}%
  %
\bibitem{Zwolak21-AAQ}%
  \BibitemOpen
  \bibfield  {author} {\bibinfo {author} {\bibfnamefont {J.~P.}\ \bibnamefont
  {Zwolak}}\ and\ \bibinfo {author} {\bibfnamefont {J.~M.}\ \bibnamefont
  {Taylor}},\ }\bibfield  {title} {\bibinfo {title} {{\it Colloquium}: Advances
  in automation of quantum dot devices control},\ }\href
  {https://doi.org/10.1103/RevModPhys.95.011006} {\bibfield  {journal}
  {\bibinfo  {journal} {Rev. Mod. Phys.}\ }\textbf {\bibinfo {volume} {95}},\
  \bibinfo {pages} {011006} (\bibinfo {year} {2023})}\BibitemShut {NoStop}%
  %
\bibitem {Moon20-ATQ}%
  \BibitemOpen
  \bibfield  {author} {\bibinfo {author} {\bibfnamefont {H.}~\bibnamefont
  {Moon}}, \bibinfo {author} {\bibfnamefont {D.~T.}\ \bibnamefont {Lennon}},
  \bibinfo {author} {\bibfnamefont {J.}~\bibnamefont {Kirkpatrick}}, \bibinfo
  {author} {\bibfnamefont {N.~M.}\ \bibnamefont {van Esbroeck}}, \bibinfo
  {author} {\bibfnamefont {L.~C.}\ \bibnamefont {Camenzind}}, \bibinfo {author}
  {\bibfnamefont {L.}~\bibnamefont {Yu}}, \bibinfo {author} {\bibfnamefont
  {F.}~\bibnamefont {Vigneau}}, \bibinfo {author} {\bibfnamefont {D.~M.}\
  \bibnamefont {Zumb\"uhl}}, \bibinfo {author} {\bibfnamefont {G.~A.~D.}\
  \bibnamefont {Briggs}}, \bibinfo {author} {\bibfnamefont {M.~A.}\
  \bibnamefont {Osborne}}, \bibinfo {author} {\bibfnamefont {D.}~\bibnamefont
  {Sejdinovic}}, \emph {et~al.},\ }{ \bibinfo {title} {Machine learning enables completely
  automatic tuning of a quantum device faster than human experts},\ }\href
  {https://doi.org/10.1038/s41467-020-17835-9} {\bibfield  {journal} {\bibinfo
  {journal} {Nat. Commun.}\ }\textbf {\bibinfo {volume} {11}},\ \bibinfo
  {pages} {4161} (\bibinfo {year} {2020})}\BibitemShut {NoStop}%
  %
\bibitem {Ziegler22-TRA}%
  \BibitemOpen
  \bibfield  {author} {\bibinfo {author} {\bibfnamefont {J.}~\bibnamefont
  {Ziegler}}, \bibinfo {author} {\bibfnamefont {T.}~\bibnamefont {McJunkin}},
  \bibinfo {author} {\bibfnamefont {E.~S.}\ \bibnamefont {Joseph}}, \bibinfo
  {author} {\bibfnamefont {S.~S.}\ \bibnamefont {Kalantre}}, \bibinfo {author}
  {\bibfnamefont {B.}~\bibnamefont {Harpt}}, \bibinfo {author} {\bibfnamefont
  {D.~E.}\ \bibnamefont {Savage}}, \bibinfo {author} {\bibfnamefont {M.~G.}\
  \bibnamefont {Lagally}}, \bibinfo {author} {\bibfnamefont {M.~A.}\
  \bibnamefont {Eriksson}}, \bibinfo {author} {\bibfnamefont {J.~M.}\
  \bibnamefont {Taylor}},\ and\ \bibinfo {author} {\bibfnamefont {J.~P.}\
  \bibnamefont {Zwolak}},\ }\bibfield  {title} {\bibinfo {title} {Toward robust
  autotuning of noisy quantum dot devices},\ }\href
  {https://doi.org/10.1103/PhysRevApplied.17.024069} {\bibfield  {journal}
  {\bibinfo  {journal} {Phys. Rev. Appl.}\ }\textbf {\bibinfo {volume} {17}},\
  \bibinfo {pages} {024069} (\bibinfo {year} {2022})}\BibitemShut {NoStop}%
  %
\bibitem {Loss98-QCD}%
  \BibitemOpen
  \bibfield  {author} {\bibinfo {author} {\bibfnamefont {D.}~\bibnamefont
  {Loss}}\ and\ \bibinfo {author} {\bibfnamefont {D.~P.}\ \bibnamefont
  {DiVincenzo}},\ }\bibfield  {title} {\bibinfo {title} {Quantum computation
  with quantum dots},\ }\href {https://doi.org/10.1103/PhysRevA.57.120}
  {\bibfield  {journal} {\bibinfo  {journal} {Phys. Rev. A}\ }\textbf {\bibinfo
  {volume} {57}},\ \bibinfo {pages} {120} (\bibinfo {year} {1998})}\BibitemShut
  {NoStop}%
  %
\bibitem {Vandersypen17-ISQ}%
  \BibitemOpen
  \bibfield  {author} {\bibinfo {author} {\bibfnamefont {L.~M.~K.}\
  \bibnamefont {Vandersypen}}, \bibinfo {author} {\bibfnamefont
  {H.}~\bibnamefont {Bluhm}}, \bibinfo {author} {\bibfnamefont {J.~S.}\
  \bibnamefont {Clarke}}, \bibinfo {author} {\bibfnamefont {A.~S.}\
  \bibnamefont {Dzurak}}, \bibinfo {author} {\bibfnamefont {R.}~\bibnamefont
  {Ishihara}}, \bibinfo {author} {\bibfnamefont {A.}~\bibnamefont {Morello}},
  \bibinfo {author} {\bibfnamefont {D.~J.}\ \bibnamefont {Reilly}}, \bibinfo
  {author} {\bibfnamefont {L.~R.}\ \bibnamefont {Schreiber}},\ and\ \bibinfo
  {author} {\bibfnamefont {M.}~\bibnamefont {Veldhorst}},\ }\bibfield  {title}
  {\bibinfo {title} {Interfacing spin qubits in quantum dots and donors—hot,
  dense, and coherent},\ }\href {https://doi.org/10.1038/s41534-017-0038-y}
  {\bibfield  {journal} {\bibinfo  {journal} {npj Quantum Inf.}\ }\textbf
  {\bibinfo {volume} {3}},\ \bibinfo {pages} {34} (\bibinfo {year}
  {2017})}\BibitemShut {NoStop}%
  %
\bibitem {Rimbach2023}%
  \BibitemOpen
  \bibfield  {author} {\bibinfo {author} {\bibfnamefont {M.}~\bibnamefont
  {Rimbach-Russ}}, \bibinfo {author} {\bibfnamefont {S.~G.~J.}\ \bibnamefont
  {Philips}}, \bibinfo {author} {\bibfnamefont {X.}~\bibnamefont {Xue}},\ and\
  \bibinfo {author} {\bibfnamefont {L.~M.~K.}\ \bibnamefont {Vandersypen}},\
  }\bibfield  {title} {\bibinfo {title} {Simple framework for systematic
  high-fidelity gate operations},\ }\href
  {https://doi.org/10.1088/2058-9565/acf786} {\bibfield  {journal} {\bibinfo
  {journal} {Quantum Sci. Technol.}\ }\textbf {\bibinfo {volume}
  {8}},\ \bibinfo {pages} {045025} (\bibinfo {year} {2023})}\BibitemShut
  {NoStop}%
  %
\bibitem {Perron15-QSB}%
  \BibitemOpen
  \bibfield  {author} {\bibinfo {author} {\bibfnamefont {J.~K.}\ \bibnamefont
  {Perron}}, \bibinfo {author} {\bibfnamefont {M.~D.}\ \bibnamefont
  {Stewart~Jr}},\ and\ \bibinfo {author} {\bibfnamefont {N.~M.}\ \bibnamefont
  {Zimmerman}},\ }\bibfield  {title} {\bibinfo {title} {A quantitative study of
  bias triangles presented in chemical potential space},\ }\href
  {https://doi.org/10.1088/0953-8984/27/23/235302} {\bibfield  {journal}
  {\bibinfo  {journal} {J. Phys.: Condens. Matter}\ }\textbf {\bibinfo {volume}
  {27}},\ \bibinfo {pages} {235302} (\bibinfo {year} {2015})}\BibitemShut
  {NoStop}%
  %
\bibitem {Hensgens17-FHQ}%
  \BibitemOpen
  \bibfield  {author} {\bibinfo {author} {\bibfnamefont {T.}~\bibnamefont
  {Hensgens}}, \bibinfo {author} {\bibfnamefont {T.}~\bibnamefont {Fujita}},
  \bibinfo {author} {\bibfnamefont {L.}~\bibnamefont {Janssen}}, \bibinfo
  {author} {\bibfnamefont {X.}~\bibnamefont {Li}}, \bibinfo {author}
  {\bibfnamefont {C.~J.}\ \bibnamefont {Van~Diepen}}, \bibinfo {author}
  {\bibfnamefont {C.}~\bibnamefont {Reichl}}, \bibinfo {author} {\bibfnamefont
  {W.}~\bibnamefont {Wegscheider}}, \bibinfo {author} {\bibfnamefont {S.~D.}\
  \bibnamefont {Sarma}},\ and\ \bibinfo {author} {\bibfnamefont {L.~M.~K.}\
  \bibnamefont {Vandersypen}},\ }\bibfield  {title} {\bibinfo {title} {Quantum
  simulation of a fermi--hubbard model using a semiconductor quantum dot
  array},\ }\href {https://doi.org/doi.org/10.1038/nature23022} {\bibfield
  {journal} {\bibinfo  {journal} {Nature}\ }\textbf {\bibinfo {volume} {548}},\
  \bibinfo {pages} {70} (\bibinfo {year} {2017})}\BibitemShut {NoStop}%
  %
\bibitem {Hensgens18-PhD}%
  \BibitemOpen
  \bibfield  {author} {\bibinfo {author} {\bibfnamefont {T.}~\bibnamefont
  {Hensgens}},\ }\emph {\bibinfo {title} {Emulating Fermi-Hubbard physics with
  quantum dots: from few to more and how to}},\ \href
  {https://doi.org/10.4233/uuid:b71f3b0b-73a0-4996-896c-84ed43e72035} {Ph.D.
  thesis},\ \bibinfo  {school} {Delft University of Technology}, \bibinfo
  {address} {Delft, Netherlands} (\bibinfo {year} {2018})\BibitemShut {NoStop}%
  %
\bibitem {Volk2019}%
  \BibitemOpen
  \bibfield  {author} {\bibinfo {author} {\bibfnamefont {C.}~\bibnamefont
  {Volk}}, \bibinfo {author} {\bibfnamefont {A.~M.~J.}\ \bibnamefont
  {Zwerver}}, \bibinfo {author} {\bibfnamefont {U.}~\bibnamefont
  {Mukhopadhyay}}, \bibinfo {author} {\bibfnamefont {P.~T.}\ \bibnamefont
  {Eendebak}}, \bibinfo {author} {\bibfnamefont {C.~J.}\ \bibnamefont {van
  Diepen}}, \bibinfo {author} {\bibfnamefont {J.~P.}\ \bibnamefont
  {Dehollain}}, \bibinfo {author} {\bibfnamefont {T.}~\bibnamefont {Hensgens}},
  \bibinfo {author} {\bibfnamefont {T.}~\bibnamefont {Fujita}}, \bibinfo
  {author} {\bibfnamefont {C.}~\bibnamefont {Reichl}}, \bibinfo {author}
  {\bibfnamefont {W.}~\bibnamefont {Wegscheider}},\ and\ \bibinfo {author}
  {\bibfnamefont {L.~M.~K.}\ \bibnamefont {Vandersypen}},\ }\bibfield  {title}
  {\bibinfo {title} {Loading a quantum-dot based ``qubyte'' register},\ }\href
  {https://doi.org/10.1038/s41534-019-0146-y} {\bibfield  {journal} {\bibinfo
  {journal} {npj Quantum Inf.}\ }\textbf {\bibinfo {volume} {5}},\
  \bibinfo {pages} {29} (\bibinfo {year} {2019})}\BibitemShut {NoStop}%
  %
\bibitem {Qiao20-CME}%
  \BibitemOpen
  \bibfield  {author} {\bibinfo {author} {\bibfnamefont {H.}~\bibnamefont
  {Qiao}}, \bibinfo {author} {\bibfnamefont {Y.~P.}\ \bibnamefont {Kandel}},
  \bibinfo {author} {\bibfnamefont {K.}~\bibnamefont {Deng}}, \bibinfo {author}
  {\bibfnamefont {S.}~\bibnamefont {Fallahi}}, \bibinfo {author} {\bibfnamefont
  {G.~C.}\ \bibnamefont {Gardner}}, \bibinfo {author} {\bibfnamefont {M.~J.}\
  \bibnamefont {Manfra}}, \bibinfo {author} {\bibfnamefont {E.}~\bibnamefont
  {Barnes}},\ and\ \bibinfo {author} {\bibfnamefont {J.~M.}\ \bibnamefont
  {Nichol}},\ }\bibfield  {title} {\bibinfo {title} {Coherent multispin
  exchange coupling in a quantum-dot spin chain},\ }\href
  {https://doi.org/10.1103/PhysRevX.10.031006} {\bibfield  {journal} {\bibinfo
  {journal} {Phys. Rev. X}\ }\textbf {\bibinfo {volume} {10}},\ \bibinfo
  {pages} {031006} (\bibinfo {year} {2020})}\BibitemShut {NoStop}%
  %
\bibitem {Hsiao20-EOT}%
  \BibitemOpen
  \bibfield  {author} {\bibinfo {author} {\bibfnamefont {T.-K.}\ \bibnamefont
  {Hsiao}}, \bibinfo {author} {\bibfnamefont {C.~J.}\ \bibnamefont {van
  Diepen}}, \bibinfo {author} {\bibfnamefont {U.}~\bibnamefont {Mukhopadhyay}},
  \bibinfo {author} {\bibfnamefont {C.}~\bibnamefont {Reichl}}, \bibinfo
  {author} {\bibfnamefont {W.}~\bibnamefont {Wegscheider}},\ and\ \bibinfo
  {author} {\bibfnamefont {L.~M.~K.}\ \bibnamefont {Vandersypen}},\ }\bibfield
  {title} {\bibinfo {title} {Efficient orthogonal control of tunnel couplings
  in a quantum dot array},\ }\href
  {https://doi.org/10.1103/PhysRevApplied.13.054018} {\bibfield  {journal}
  {\bibinfo  {journal} {Phys. Rev. Appl.}\ }\textbf {\bibinfo {volume} {13}},\
  \bibinfo {pages} {054018} (\bibinfo {year} {2020})}\BibitemShut {NoStop}%
  %
\bibitem {Ziegler23-AEC}%
  \BibitemOpen
  \bibfield  {author} {\bibinfo {author} {\bibfnamefont {J.}~\bibnamefont
  {Ziegler}}, \bibinfo {author} {\bibfnamefont {F.}~\bibnamefont {Luthi}},
  \bibinfo {author} {\bibfnamefont {M.}~\bibnamefont {Ramsey}}, \bibinfo
  {author} {\bibfnamefont {F.}~\bibnamefont {Borjans}}, \bibinfo {author}
  {\bibfnamefont {G.}~\bibnamefont {Zheng}},\ and\ \bibinfo {author}
  {\bibfnamefont {J.~P.}\ \bibnamefont {Zwolak}},\ }\bibfield  {title}
  {\bibinfo {title} {Automated extraction of capacitive coupling for quantum
  dot systems},\ }\href {https://doi.org/10.1103/PhysRevApplied.19.054077}
  {\bibfield  {journal} {\bibinfo  {journal} {Phys. Rev. Appl.}\ }\textbf
  {\bibinfo {volume} {19}},\ \bibinfo {pages} {054077} (\bibinfo {year}
  {2023}{\natexlab{b}})}\BibitemShut {NoStop}%
  %
\bibitem {Oakes2021automatic}%
  \BibitemOpen
  \bibfield  {author} {\bibinfo {author} {\bibfnamefont {G.~A.}\ \bibnamefont
  {Oakes}}, \bibinfo {author} {\bibfnamefont {J.}~\bibnamefont {Duan}},
  \bibinfo {author} {\bibfnamefont {J.~J.~L.}\ \bibnamefont {Morton}}, \bibinfo
  {author} {\bibfnamefont {A.}~\bibnamefont {Lee}}, \bibinfo {author}
  {\bibfnamefont {C.~G.}\ \bibnamefont {Smith}},\ and\ \bibinfo {author}
  {\bibfnamefont {M.~F.~G.}\ \bibnamefont {Zalba}},\ }{\bibinfo {title} {Automatic virtual
  voltage extraction of a 2x2 array of quantum dots with machine learning}},\ 
  \Eprint {https://arxiv.org/abs/2012.03685}
  {arXiv:2012.03685}  (\bibinfo {year} {2021})\BibitemShut {NoStop}%
  %
\bibitem {Mills19-CAT}%
  \BibitemOpen
  \bibfield  {author} {\bibinfo {author} {\bibfnamefont {A.~R.}\ \bibnamefont
  {Mills}}, \bibinfo {author} {\bibfnamefont {M.~M.}\ \bibnamefont {Feldman}},
  \bibinfo {author} {\bibfnamefont {C.}~\bibnamefont {Monical}}, \bibinfo
  {author} {\bibfnamefont {P.~J.}\ \bibnamefont {Lewis}}, \bibinfo {author}
  {\bibfnamefont {K.~W.}\ \bibnamefont {Larson}}, \bibinfo {author}
  {\bibfnamefont {A.~M.}\ \bibnamefont {Mounce}},\ and\ \bibinfo {author}
  {\bibfnamefont {J.~R.}\ \bibnamefont {Petta}},\ }\bibfield  {title} {\bibinfo
  {title} {Computer-automated tuning procedures for semiconductor quantum dot
  arrays},\ }\href {https://doi.org/10.1063/1.5121444} {\bibfield  {journal}
  {\bibinfo  {journal} {Appl. Phys. Lett.}\ }\textbf {\bibinfo {volume}
  {115}},\ \bibinfo {pages} {113501} (\bibinfo {year} {2019})}\BibitemShut
  {NoStop}%
  %
\bibitem {Lapointe-Major19-ATQ}%
  \BibitemOpen
  \bibfield  {author} {\bibinfo {author} {\bibfnamefont {M.}~\bibnamefont
  {Lapointe-Major}}, \bibinfo {author} {\bibfnamefont {O.}~\bibnamefont
  {Germain}}, \bibinfo {author} {\bibfnamefont {J.}~\bibnamefont
  {Camirand~Lemyre}}, \bibinfo {author} {\bibfnamefont {D.}~\bibnamefont
  {Lachance-Quirion}}, \bibinfo {author} {\bibfnamefont {S.}~\bibnamefont
  {Rochette}}, \bibinfo {author} {\bibfnamefont {F.}~\bibnamefont
  {Camirand~Lemyre}},\ and\ \bibinfo {author} {\bibfnamefont {M.}~\bibnamefont
  {Pioro-Ladri\`ere}},\ }\bibfield  {title} {\bibinfo {title} {Algorithm for
  automated tuning of a quantum dot into the single-electron regime},\ }\href
  {https://doi.org/10.1103/PhysRevB.102.085301} {\bibfield  {journal} {\bibinfo
   {journal} {Phys. Rev. B}\ }\textbf {\bibinfo {volume} {102}},\ \bibinfo
  {pages} {085301} (\bibinfo {year} {2020})}\BibitemShut {NoStop}%
  %
\bibitem {Oosterkamp98-MSQ}%
  \BibitemOpen
  \bibfield  {author} {\bibinfo {author} {\bibfnamefont {T.~H.}\ \bibnamefont
  {Oosterkamp}}, \bibinfo {author} {\bibfnamefont {T.}~\bibnamefont
  {Fujisawa}}, \bibinfo {author} {\bibfnamefont {W.~G.}\ \bibnamefont {van~der
  Wiel}}, \bibinfo {author} {\bibfnamefont {K.}~\bibnamefont {Ishibashi}},
  \bibinfo {author} {\bibfnamefont {R.~V.}\ \bibnamefont {Hijman}}, \bibinfo
  {author} {\bibfnamefont {S.}~\bibnamefont {Tarucha}},\ and\ \bibinfo {author}
  {\bibfnamefont {L.~P.}\ \bibnamefont {Kouwenhoven}},\ }\bibfield  {title}
  {\bibinfo {title} {Microwave spectroscopy of a quantum-dot molecule},\ }\href
  {https://doi.org/10.1038/27617} {\bibfield  {journal} {\bibinfo  {journal}
  {Nature}\ }\textbf {\bibinfo {volume} {395}},\ \bibinfo {pages} {873}
  (\bibinfo {year} {1998})}\BibitemShut {NoStop}%
  %
\bibitem {Stehouwer24-ESG}%
 \BibitemOpen
 \bibfield  {author} {\bibinfo {author} {\bibfnamefont {L.~E.~A.}\
  \bibnamefont {Stehouwer}}, \bibinfo {author} {\bibfnamefont {C.~X.}\
  \bibnamefont {Yu}}, \bibinfo {author} {\bibfnamefont {B.}~\bibnamefont {van
  Straaten}}, \bibinfo {author} {\bibfnamefont {A.}~\bibnamefont {Tosato}},
  \bibinfo {author} {\bibfnamefont {V.}~\bibnamefont {John}}, \bibinfo {author}
  {\bibfnamefont {D.~D.}\ \bibnamefont {Esposti}}, \bibinfo {author}
  {\bibfnamefont {A.}~\bibnamefont {Elsayed}}, \bibinfo {author} {\bibfnamefont
  {D.}~\bibnamefont {Costa}}, \bibinfo {author} {\bibfnamefont {S.~D.}\
  \bibnamefont {Oosterhout}}, \bibinfo {author} {\bibfnamefont {N.~W.}\
  \bibnamefont {Hendrickx}}, \emph {et~al.},\ }{\bibinfo {title} {Exploiting epitaxial
  strained germanium for scaling low noise spin qubits at the micron-scale}},\ 
  \Eprint {https://arxiv.org/abs/2411.11526}{arXiv:2411.11526} 
  (\bibinfo {year} {2024})\BibitemShut {NoStop}%
  %
\bibitem {John24-TAL}%
  \BibitemOpen
  \bibfield  {author} {\bibinfo {author} {\bibfnamefont {V.}~\bibnamefont
  {John}}, \bibinfo {author} {\bibfnamefont {C.~X.}\ \bibnamefont {Yu}},
  \bibinfo {author} {\bibfnamefont {B.}~\bibnamefont {van Straaten}}, \bibinfo
  {author} {\bibfnamefont {E.~A.}\ \bibnamefont {Rodríguez-Mena}}, \bibinfo
  {author} {\bibfnamefont {M.}~\bibnamefont {Rodríguez}}, \bibinfo {author}
  {\bibfnamefont {S.}~\bibnamefont {Oosterhout}}, \bibinfo {author}
  {\bibfnamefont {L.~E.~A.}\ \bibnamefont {Stehouwer}}, \bibinfo {author}
  {\bibfnamefont {G.}~\bibnamefont {Scappucci}}, \bibinfo {author}
  {\bibfnamefont {S.}~\bibnamefont {Bosco}}, \bibinfo {author} {\bibfnamefont
  {M.}~\bibnamefont {Rimbach-Russ}}, \emph {et~al.},\ }\bibfield  {title} {\bibinfo {title} {A two-dimensional
  10-qubit array in germanium with robust and localised qubit control},\ }\href
  {https://arxiv.org/abs/2412.16044} {\bibfield  {journal} {\bibinfo  {journal}
  {arXiv:2412.16044}\ } (\bibinfo {year} {2024})}\BibitemShut {NoStop}%
  %
\bibitem {stehlik2015}%
  \BibitemOpen
  \bibfield  {author} {\bibinfo {author} {\bibfnamefont {J.}~\bibnamefont
  {Stehlik}}, \bibinfo {author} {\bibfnamefont {Y.-Y.}\ \bibnamefont {Liu}},
  \bibinfo {author} {\bibfnamefont {C.~M.}\ \bibnamefont {Quintana}}, \bibinfo
  {author} {\bibfnamefont {C.}~\bibnamefont {Eichler}}, \bibinfo {author}
  {\bibfnamefont {T.~R.}\ \bibnamefont {Hartke}},\ and\ \bibinfo {author}
  {\bibfnamefont {J.~R.}\ \bibnamefont {Petta}},\ }\bibfield  {title} {\bibinfo
  {title} {Fast charge sensing of a cavity-coupled double quantum dot using a
  josephson parametric amplifier},\ }\href
  {https://doi.org/10.1103/PhysRevApplied.4.014018} {\bibfield  {journal}
  {\bibinfo  {journal} {Phys. Rev. Appl.}\ }\textbf {\bibinfo {volume} {4}},\
  \bibinfo {pages} {014018} (\bibinfo {year} {2015})}\BibitemShut {NoStop}%
  %
\bibitem {vigneau2023}%
  \BibitemOpen
  \bibfield  {author} {\bibinfo {author} {\bibfnamefont {F.}~\bibnamefont
  {Vigneau}}, \bibinfo {author} {\bibfnamefont {F.}~\bibnamefont {Fedele}},
  \bibinfo {author} {\bibfnamefont {A.}~\bibnamefont {Chatterjee}}, \bibinfo
  {author} {\bibfnamefont {D.}~\bibnamefont {Reilly}}, \bibinfo {author}
  {\bibfnamefont {F.}~\bibnamefont {Kuemmeth}}, \bibinfo {author}
  {\bibfnamefont {M.~F.}\ \bibnamefont {Gonzalez-Zalba}}, \bibinfo {author}
  {\bibfnamefont {E.}~\bibnamefont {Laird}},\ and\ \bibinfo {author}
  {\bibfnamefont {N.}~\bibnamefont {Ares}},\ }\bibfield  {title} {\bibinfo
  {title} {{Probing quantum devices with radio-frequency reflectometry}},\
  }\href {https://doi.org/10.1063/5.0088229} {\bibfield  {journal} {\bibinfo
  {journal} {Appl. Phys. Rev.}\ }\textbf {\bibinfo {volume} {10}},\
  \bibinfo {pages} {021305} (\bibinfo {year} {2023})}\BibitemShut {NoStop}%
  %
\bibitem {Chollet20-ISU}%
  \BibitemOpen
  \bibfield  {author} {\bibinfo {author} {\bibfnamefont {F.}~\bibnamefont
  {Chollet}},\ }\href
  {https://keras.io/examples/vision/oxford_pets_image_segmentation/} {\bibinfo
  {title} {Image segmentation with a U-Net-like architecture}} (\bibinfo {year}
  {2019})\BibitemShut {NoStop}%
  %
\bibitem {Zwolak18-QLD}%
  \BibitemOpen
  \bibfield  {author} {\bibinfo {author} {\bibfnamefont {J.~P.}\ \bibnamefont
  {Zwolak}}, \bibinfo {author} {\bibfnamefont {S.~S.}\ \bibnamefont
  {Kalantre}}, \bibinfo {author} {\bibfnamefont {X.}~\bibnamefont {Wu}},
  \bibinfo {author} {\bibfnamefont {S.}~\bibnamefont {Ragole}},\ and\ \bibinfo
  {author} {\bibfnamefont {J.~M.}\ \bibnamefont {Taylor}},\ }\bibfield  {title}
  {\bibinfo {title} {{QFlow} lite dataset: {A} machine-learning approach to the
  charge states in quantum dot experiments},\ }\href
  {https://doi.org/10.1371/journal.pone.0205844} {\bibfield  {journal}
  {\bibinfo  {journal} {PLoS ONE}\ }\textbf {\bibinfo {volume} {13}},\ \bibinfo
  {pages} {e0205844} (\bibinfo {year} {2018})}\BibitemShut {NoStop}%
  %
\bibitem {qf-data}%
  \BibitemOpen
  \bibfield  {author} {\bibinfo {author} {\bibnamefont {{National Institute of
  Standards and Technology}}},\ }\href@noop {} {\bibinfo {title} {Qflow 2.0:
  Quantum dot data for machine learning}},\ \bibinfo {howpublished} {Database:
  data.nist.gov, \url{https://doi.org/10.18434/T4/1423788}} (\bibinfo {year}
  {2022})\BibitemShut {NoStop}%
  %
\bibitem {Giusti13-FIS}%
  \BibitemOpen
  \bibfield  {author} {\bibinfo {author} {\bibfnamefont {A.}~\bibnamefont
  {Giusti}}, \bibinfo {author} {\bibfnamefont {D.~C.}\ \bibnamefont
  {Cire{\c{s}}an}}, \bibinfo {author} {\bibfnamefont {J.}~\bibnamefont
  {Masci}}, \bibinfo {author} {\bibfnamefont {L.~M.}\ \bibnamefont
  {Gambardella}},\ and\ \bibinfo {author} {\bibfnamefont {J.}~\bibnamefont
  {Schmidhuber}},\ }\bibfield  {title} {\bibinfo {title} {Fast image scanning
  with deep max-pooling convolutional neural networks},\ }in\ \href@noop {}
  {\emph {\bibinfo {booktitle} {2013 IEEE international conference on image
  processing}}}\ (\bibinfo {organization} {IEEE},\ \bibinfo {year} {2013})\
  pp.\ \bibinfo {pages} {4034--4038}\BibitemShut {NoStop}%
  %
\bibitem {Gouk14-FSW}%
  \BibitemOpen
  \bibfield  {author} {\bibinfo {author} {\bibfnamefont {H.~G.}\ \bibnamefont
  {Gouk}}\ and\ \bibinfo {author} {\bibfnamefont {A.~M.}\ \bibnamefont
  {Blake}},\ }\bibfield  {title} {\bibinfo {title} {Fast sliding window
  classification with convolutional neural networks},\ }in\ \href@noop {}
  {\emph {\bibinfo {booktitle} {Proceedings of the 29th International
  Conference on Image and Vision Computing New Zealand}}}\ (\bibinfo {year}
  {2014})\ pp.\ \bibinfo {pages} {114--118}\BibitemShut {NoStop}%
  %
\bibitem {Note1}%
  \BibitemOpen
  \bibinfo {note} {By design, the pixel classifiers return five-class output.
  However, the diagonal transition and no transition classes are not relevant
  to the analysis.}\BibitemShut {Stop}%
  %
\bibitem {supp}%
  \BibitemOpen
  \href@noop {} {}\bibinfo {note} {See Supplemental Material for
  discussion of MAViS scalability and complexity as well as additional
  figures.}\BibitemShut {Stop}%
  %
\bibitem {Dvir2023}%
  \BibitemOpen
  \bibfield  {author} {\bibinfo {author} {\bibfnamefont {T.}~\bibnamefont
  {Dvir}}, \bibinfo {author} {\bibfnamefont {G.}~\bibnamefont {Wang}}, \bibinfo
  {author} {\bibfnamefont {N.}~\bibnamefont {van Loo}}, \bibinfo {author}
  {\bibfnamefont {C.-X.}\ \bibnamefont {Liu}}, \bibinfo {author} {\bibfnamefont
  {G.~P.}\ \bibnamefont {Mazur}}, \bibinfo {author} {\bibfnamefont
  {A.}~\bibnamefont {Bordin}}, \bibinfo {author} {\bibfnamefont {S.~L.~D.}\
  \bibnamefont {ten Haaf}}, \bibinfo {author} {\bibfnamefont {J.-Y.}\
  \bibnamefont {Wang}}, \bibinfo {author} {\bibfnamefont {D.}~\bibnamefont {van
  Driel}}, \bibinfo {author} {\bibfnamefont {F.}~\bibnamefont {Zatelli}}, \emph {et~al.},\ }\bibfield  {title} {\bibinfo {title}
  {Realization of a minimal kitaev chain in coupled quantum dots},\ }\href
  {https://doi.org/10.1038/s41586-022-05585-1} {\bibfield  {journal} {\bibinfo
  {journal} {Nature}\ }\textbf {\bibinfo {volume} {614}},\ \bibinfo {pages}
  {445} (\bibinfo {year} {2023})}\BibitemShut {NoStop}%
  %
\bibitem {tenHaaf2024}%
  \BibitemOpen
  \bibfield  {author} {\bibinfo {author} {\bibfnamefont {S.~L.~D.}\
  \bibnamefont {ten Haaf}}, \bibinfo {author} {\bibfnamefont {Q.}~\bibnamefont
  {Wang}}, \bibinfo {author} {\bibfnamefont {A.~M.}\ \bibnamefont {Bozkurt}},
  \bibinfo {author} {\bibfnamefont {C.-X.}\ \bibnamefont {Liu}}, \bibinfo
  {author} {\bibfnamefont {I.}~\bibnamefont {Kulesh}}, \bibinfo {author}
  {\bibfnamefont {P.}~\bibnamefont {Kim}}, \bibinfo {author} {\bibfnamefont
  {D.}~\bibnamefont {Xiao}}, \bibinfo {author} {\bibfnamefont {C.}~\bibnamefont
  {Thomas}}, \bibinfo {author} {\bibfnamefont {M.~J.}\ \bibnamefont {Manfra}},
  \bibinfo {author} {\bibfnamefont {T.}~\bibnamefont {Dvir}}, \emph {et~al.},\ }\bibfield  {title} {\bibinfo
  {title} {A two-site kitaev chain in a two-dimensional electron gas},\ }\href
  {https://doi.org/10.1038/s41586-024-07434-9} {\bibfield  {journal} {\bibinfo
  {journal} {Nature}\ }\textbf {\bibinfo {volume} {630}},\ \bibinfo {pages}
  {329} (\bibinfo {year} {2024})}\BibitemShut {NoStop}%
  %
\bibitem {Bordin2024}%
  \BibitemOpen
  \bibfield  {author} {\bibinfo {author} {\bibfnamefont {A.}~\bibnamefont
  {Bordin}}, \bibinfo {author} {\bibfnamefont {C.-X.}\ \bibnamefont {Liu}},
  \bibinfo {author} {\bibfnamefont {T.}~\bibnamefont {Dvir}}, \bibinfo {author}
  {\bibfnamefont {F.}~\bibnamefont {Zatelli}}, \bibinfo {author} {\bibfnamefont
  {S.~L.~D.}\ \bibnamefont {ten Haaf}}, \bibinfo {author} {\bibfnamefont
  {D.}~\bibnamefont {van Driel}}, \bibinfo {author} {\bibfnamefont
  {G.}~\bibnamefont {Wang}}, \bibinfo {author} {\bibfnamefont {N.}~\bibnamefont
  {van Loo}}, \bibinfo {author} {\bibfnamefont {T.}~\bibnamefont {van
  Caekenberghe}}, \bibinfo {author} {\bibfnamefont {J.~C.}\ \bibnamefont
  {Wolff}}, \emph {et~al.},\ }{\bibinfo {title} {Signatures of majorana protection in a three-site kitaev
  chain}},\ \Eprint {https://arxiv.org/abs/2402.19382}
  {arXiv:2402.19382} (\bibinfo {year} {2024})\BibitemShut {NoStop}%
  %
\bibitem {Hays2021}%
  \BibitemOpen
  \bibfield  {author} {\bibinfo {author} {\bibfnamefont {M.}~\bibnamefont
  {Hays}}, \bibinfo {author} {\bibfnamefont {V.}~\bibnamefont {Fatemi}},
  \bibinfo {author} {\bibfnamefont {D.}~\bibnamefont {Bouman}}, \bibinfo
  {author} {\bibfnamefont {J.}~\bibnamefont {Cerrillo}}, \bibinfo {author}
  {\bibfnamefont {S.}~\bibnamefont {Diamond}}, \bibinfo {author} {\bibfnamefont
  {K.}~\bibnamefont {Serniak}}, \bibinfo {author} {\bibfnamefont
  {T.}~\bibnamefont {Connolly}}, \bibinfo {author} {\bibfnamefont
  {P.}~\bibnamefont {Krogstrup}}, \bibinfo {author} {\bibfnamefont
  {J.}~\bibnamefont {Nygård}}, \bibinfo {author} {\bibfnamefont {A.~L.}\
  \bibnamefont {Yeyati}}, \emph {et~al.},\ }\bibfield  {title} {\bibinfo {title} {Coherent manipulation of
  an andreev spin qubit},\ }\href {https://doi.org/10.1126/science.abf0345}
  {\bibfield  {journal} {\bibinfo  {journal} {Science}\ }\textbf {\bibinfo
  {volume} {373}},\ \bibinfo {pages} {430} (\bibinfo {year} {2021})} \BibitemShut
  {NoStop}%
  %
\bibitem {Pita-Vidal2023}%
  \BibitemOpen
  \bibfield  {author} {\bibinfo {author} {\bibfnamefont {M.}~\bibnamefont
  {Pita-Vidal}}, \bibinfo {author} {\bibfnamefont {A.}~\bibnamefont
  {Bargerbos}}, \bibinfo {author} {\bibfnamefont {R.}~\bibnamefont
  {{\v{Z}}itko}}, \bibinfo {author} {\bibfnamefont {L.~J.}\ \bibnamefont
  {Splitthoff}}, \bibinfo {author} {\bibfnamefont {L.}~\bibnamefont
  {Gr{\"u}nhaupt}}, \bibinfo {author} {\bibfnamefont {J.~J.}\ \bibnamefont
  {Wesdorp}}, \bibinfo {author} {\bibfnamefont {Y.}~\bibnamefont {Liu}},
  \bibinfo {author} {\bibfnamefont {L.~P.}\ \bibnamefont {Kouwenhoven}},
  \bibinfo {author} {\bibfnamefont {R.}~\bibnamefont {Aguado}}, \bibinfo
  {author} {\bibfnamefont {B.}~\bibnamefont {van Heck}}, \emph {et~al.},\ }\bibfield  {title}
  {\bibinfo {title} {Direct manipulation of a superconducting spin qubit
  strongly coupled to a transmon qubit},\ }\href
  {https://doi.org/10.1038/s41567-023-02071-x} {\bibfield  {journal} {\bibinfo
  {journal} {Nat. Phys.}\ }\textbf {\bibinfo {volume} {19}},\ \bibinfo
  {pages} {1110} (\bibinfo {year} {2023})}\BibitemShut {NoStop}%
  %
\bibitem {Borsoi2020}%
  \BibitemOpen
  \bibfield  {author} {\bibinfo {author} {\bibfnamefont {F.}~\bibnamefont
  {Borsoi}}, \bibinfo {author} {\bibfnamefont {K.}~\bibnamefont {Zuo}},
  \bibinfo {author} {\bibfnamefont {S.}~\bibnamefont {Gazibegovic}}, \bibinfo
  {author} {\bibfnamefont {R.~L.~M.}\ \bibnamefont {Op~het Veld}}, \bibinfo
  {author} {\bibfnamefont {E.~P. A.~M.}\ \bibnamefont {Bakkers}}, \bibinfo
  {author} {\bibfnamefont {L.~P.}\ \bibnamefont {Kouwenhoven}},\ and\ \bibinfo
  {author} {\bibfnamefont {S.}~\bibnamefont {Heedt}},\ }\bibfield  {title}
  {\bibinfo {title} {Transmission phase read-out of a large quantum dot in a
  nanowire interferometer},\ }\href
  {https://doi.org/10.1038/s41467-020-17461-5} {\bibfield  {journal} {\bibinfo
  {journal} {Nat. Commun.}\ }\textbf {\bibinfo {volume} {11}},\
  \bibinfo {pages} {3666} (\bibinfo {year} {2020})}\BibitemShut {NoStop}%
  %
\bibitem {Aghaee2024}%
  \BibitemOpen
  \bibfield  {author} {\bibinfo {author} {\bibfnamefont {M.}~\bibnamefont
  {Aghaee}}, \bibinfo {author} {\bibfnamefont {A.~A.}\ \bibnamefont {Ramirez}},
  \bibinfo {author} {\bibfnamefont {Z.}~\bibnamefont {Alam}}, \bibinfo {author}
  {\bibfnamefont {R.}~\bibnamefont {Ali}}, \bibinfo {author} {\bibfnamefont
  {M.}~\bibnamefont {Andrzejczuk}}, \bibinfo {author} {\bibfnamefont
  {A.}~\bibnamefont {Antipov}}, \bibinfo {author} {\bibfnamefont
  {M.}~\bibnamefont {Astafev}}, \bibinfo {author} {\bibfnamefont
  {A.}~\bibnamefont {Barzegar}}, \bibinfo {author} {\bibfnamefont
  {B.}~\bibnamefont {Bauer}}, \bibinfo {author} {\bibfnamefont
  {J.}~\bibnamefont {Becker}}, \emph {et~al.},\ }{\bibinfo {title} {Interferometric
  single-shot parity measurement in an inas-al hybrid device}},\ \Eprint {https://arxiv.org/abs/2401.09549} {arXiv:2401.09549
 } (\bibinfo {year} {2024})\BibitemShut {NoStop}%
 %
\bibitem {Hendrickx20-FTL}%
  \BibitemOpen
  \bibfield  {author} {\bibinfo {author} {\bibfnamefont {N.~W.}\ \bibnamefont
  {Hendrickx}}, \bibinfo {author} {\bibfnamefont {D.~P.}\ \bibnamefont
  {Franke}}, \bibinfo {author} {\bibfnamefont {A.}~\bibnamefont {Sammak}},
  \bibinfo {author} {\bibfnamefont {G.}~\bibnamefont {Scappucci}},\ and\
  \bibinfo {author} {\bibfnamefont {M.}~\bibnamefont {Veldhorst}},\ }\bibfield
  {title} {\bibinfo {title} {Fast two-qubit logic with holes in germanium},\
  }\href {https://doi.org/10.1038/s41586-019-1919-3} {\bibfield  {journal}
  {\bibinfo  {journal} {Nature}\ }\textbf {\bibinfo {volume} {577}},\ \bibinfo
  {pages} {487} (\bibinfo {year} {2020})}\BibitemShut {NoStop}%
  %
\bibitem {virtualization-dataset}%
  \BibitemOpen
  \bibfield  {author} {\bibinfo {author} {\bibfnamefont {A.~S.}\ \bibnamefont
  {Rao}}, \bibinfo {author} {\bibfnamefont {D.}~\bibnamefont {Buterakos}},
  \bibinfo {author} {\bibfnamefont {B.}~\bibnamefont {van Straaten}}, \bibinfo
  {author} {\bibfnamefont {V.}~\bibnamefont {John}}, \bibinfo {author}
  {\bibfnamefont {C.~X.}\ \bibnamefont {Yu}}, \bibinfo {author} {\bibfnamefont
  {S.~D.}\ \bibnamefont {Oosterhout}}, \bibinfo {author} {\bibfnamefont
  {L.}~\bibnamefont {Stehouwer}}, \bibinfo {author} {\bibfnamefont
  {G.}~\bibnamefont {Scappucci}}, \bibinfo {author} {\bibfnamefont
  {M.}~\bibnamefont {Veldhorst}}, \bibinfo {author} {\bibfnamefont
  {F.}~\bibnamefont {Borsoi}},\ and\ \bibinfo {author} {\bibfnamefont {J.~P.}\
  \bibnamefont {Zwolak}},\ }\href@noop {} {\bibinfo {title} {Dataset underlying
  the manuscript: MAViS: Modular Autonomous Virtualization System for
  two-dimensional semiconductor quantum dot arrays}},\ \bibinfo {howpublished}
  {Zenodo: \url{https://doi.org/10.5281/zenodo.14173838}} (\bibinfo {year}
  {2024})\BibitemShut {NoStop}%
  %
\bibitem {mavis-figures}%
  \BibitemOpen
  \bibfield  {author} {\bibinfo {author} {\bibfnamefont {A.~S.}\ \bibnamefont
  {Rao}}, \bibinfo {author} {\bibfnamefont {D.}~\bibnamefont {Buterakos}},
  \bibinfo {author} {\bibfnamefont {B.}~\bibnamefont {van Straaten}}, \bibinfo
  {author} {\bibfnamefont {V.}~\bibnamefont {John}}, \bibinfo {author}
  {\bibfnamefont {C.~X.}\ \bibnamefont {Yu}}, \bibinfo {author} {\bibfnamefont
  {S.~D.}\ \bibnamefont {Oosterhout}}, \bibinfo {author} {\bibfnamefont
  {L.}~\bibnamefont {Stehouwer}}, \bibinfo {author} {\bibfnamefont
  {G.}~\bibnamefont {Scappucci}}, \bibinfo {author} {\bibfnamefont
  {M.}~\bibnamefont {Veldhorst}}, \bibinfo {author} {\bibfnamefont
  {F.}~\bibnamefont {Borsoi}},\ and\ \bibinfo {author} {\bibfnamefont {J.~P.}\
  \bibnamefont {Zwolak}},\ }\href@noop {} {\bibinfo {title} {{Figure files for
  "Modular Autonomous Virtualization System for Two-Dimensional Semiconductor
  Quantum Dot Arrays" published in Physical Review X}}},\ \bibinfo
  {howpublished} {National Institute of Standards and Technology,
  data.nist.gov: \url{https://doi.org/10.18434/mds2-3705}} (\bibinfo {year}
  {2025})\BibitemShut {NoStop}%
  %
\bibitem {vanstraaten2024}%
  \BibitemOpen
  \bibfield  {author} {\bibinfo {author} {\bibfnamefont {B.}~\bibnamefont {van
  Straaten}}, \bibinfo {author} {\bibfnamefont {J.}~\bibnamefont {Hickie}},
  \bibinfo {author} {\bibfnamefont {L.}~\bibnamefont {Schorling}}, \bibinfo
  {author} {\bibfnamefont {J.}~\bibnamefont {Schuff}}, \bibinfo {author}
  {\bibfnamefont {F.}~\bibnamefont {Fedele}},\ and\ \bibinfo {author}
  {\bibfnamefont {N.}~\bibnamefont {Ares}},\ }\bibfield  {title} {\bibinfo
  {title} {{QArray: A GPU-accelerated constant capacitance model simulator for
  large quantum dot arrays}},\ }\href
  {https://doi.org/10.21468/SciPostPhysCodeb.35} {\bibfield  {journal}
  {\bibinfo  {journal} {SciPost Phys. Codebases}\,\ \textbf{\bibinfo {pages} {35}}}
  (\bibinfo {year} {2024})}\BibitemShut {NoStop}%
\end{thebibliography}
\end{document}